\documentclass[numberedappendix]{emulateapj}
\usepackage{amstext}
\usepackage{apjfonts}
\usepackage{amsmath}
\usepackage{amssymb}
\usepackage[colorlinks=true,linkcolor=blue,citecolor=blue]{hyperref}
\usepackage{comment}
\usepackage{mathrsfs}

\newcommand{\beq}{\begin{equation}}
\newcommand{\eeq}{\end{equation}}

\begin{document}

\title{Dissipative Evolution of  Unequal Mass Binary-Single Interactions and its Relevance to Gravitational Wave  Detections} 
\author{Johan Samsing$^{1,*}$, Morgan MacLeod$^{2,*}$, Enrico Ramirez-Ruiz$^{3,4}$} 
\altaffiltext{1}{Department of Astrophysical Sciences, Princeton University, Peyton Hall, 4 Ivy Lane, Princeton, NJ 08544, USA}
\altaffiltext{2}{School of Natural Sciences, Institute for Advanced Study, 1 Einstein Drive, Princeton, New Jersey 08540, USA}
\altaffiltext{3}{Department of Astronomy and Astrophysics, University of California, Santa Cruz, CA 95064, USA}
\altaffiltext{4}{Niels Bohr Institute, University of Copenhagen, Blegdamsvej 17, 2100 Copenhagen, Denmark}

\altaffiltext{*}{Einstein Fellow}

\begin{abstract} 
We present a study on binary-single interactions with energy loss terms  such as tidal  dissipation and gravitational wave emission added to the equation-of-motion.
The inclusion of such terms leads to the formation of compact binaries that form
during the three-body interaction through two-body captures. These binaries predominantly merge relative promptly at high eccentricity,
with several observable and dynamical consequences to follow. Despite their possibility for being observed in both present and upcoming transient
surveys, their outcomes are not firmly constrained. In this paper we present an analytical framework that allows to estimate the cross section of such two-body captures,
which permits us  to study how the corresponding rates depends on the initial orbital parameters, the mass hierarchy, the type of interacting
objects, and the energy dissipation mechanism.  This formalism is applied here to study the formation of two-body gravitational wave captures, for which
we estimate absolute and relative rates relevant to Advanced LIGO detections.  It is shown that two-body gravitational wave captures should have compelling  observational implications if a sizable fraction of detected compact binaries are formed  via dynamical interactions.
\end{abstract}

\section{Introduction}
With the recent detection of gravitational waves (GWs) by LIGO \citep{2016PhRvL.116x1103A, 2016PhRvL.116f1102A, PhysRevLett.118.221101}, and upcoming
electromagnetic (EM) surveys including
LSST \citep{2009arXiv0912.0201L}, JWST \citep{2006SSRv..123..485G} and WFIRST \citep{2013arXiv1305.5422S}, a new exciting era in transient astrophysics is eminent.
To learn from the population of  current detections, that for LIGO sources can
be up to 250~Gpc$^{-3}$~yr$^{-1}$ \citep{2016arXiv160203842T, 2016ApJ...819..108B, 2016MNRAS.460.3545D}, 
major effort is now being made to model and understand the range of possible avenues such transient sources can be assembled  in the Universe \citep[e.g.][]{2017arXiv170407379Z}.

One viable avenue to assemble GW sources is through few-body dynamics, where exotic outcomes can form
as a result of chaotic or secular exchange of energy and angular momentum.
However, not all few-body interactions are equally probable; in high stellar density environments, such as a globular cluster (GC), one finds the
most frequent few-body encounters to be binary-single interactions \citep[e.g.][]{Heggie:1975uy, Hut:1983js, Hut:1983by, Hut:1993gs}. Such three-body interactions are not only important dynamically \citep{Heggie:1975uy}, but have also been shown to potentially play a role for EM transients
\citep[e.g.][]{1985ApJ...298..502H, 1986ApJ...306..552M, Fregeau:2004fj, 2007ASPC..367..697M, Rosswog:2009wq,
 Lee:2010ina, 2013MNRAS.429.2298M, 2014ApJ...794....7M, 2015ApJ...802L..22R, 2016ApJ...819...70M, 2016arXiv160909114S, 2016ApJ...823..113P}, as well
as to the assembly of binary black hole (BBH) mergers \citep{2006ApJ...640..156G, 2014MNRAS.441.3703Z, 2014ApJ...784...71S, 2015MNRAS.451.4086S,
2015PhRvL.115e1101R, 2016MNRAS.463.2443K, 2016arXiv160909114S, 2016PhRvD..93h4029R, 2016ApJ...824L...8R, 2017arXiv170309703S}. 

To understand the evolution of dynamical systems, many $N$-body simulations have been performed, from detailed few-body
studies  \citep[e.g.][]{Hut:1983js, 2014ApJ...784...71S, 2016MNRAS.456.4219A,  2016arXiv160909114S}, to full
GC simulations \citep[e.g.][]{2016MNRAS.458.1450W}. Such $N$-body simulations are often evolved using an
equation-of-motion (EOM) without terms correcting for possible orbital energy losses and finite size effects.
However, recent work have shown that a consistent inclusion of such terms surprisingly seems to play a major role for estimating accurate rates of transients and their
corresponding distribution of orbital parameters at merger \citep[e.g.][]{2006ApJ...640..156G, 2014ApJ...784...71S, 2016arXiv160909114S, 2017arXiv170309703S}.

In this paper we continue our study of binary-single interactions
with correction terms that include finite sizes, dynamical tides, and GW emission. 
The main effect from the inclusion of  such terms in the EOM is the formation of high eccentricity compact binaries that assemble
during  three-body interactions \citep{2006ApJ...640..156G, 2014ApJ...784...71S, 2016arXiv160909114S, 2017arXiv170309703S}. In order to study the relative importance of this highly eccentric merger population, this paper develops an analytical framework that allows us to robustly estimate their cross section
and corresponding rates. This formalism can be used for merging binaries driven by  tidal  dissipation or  gravitational wave emission.

We apply our analytical framework to study the assembly of BBH mergers forming through binary-single interactions
as a result of GW emission and its dependence on the initial orbital parameters and the mass hierarchy.  This exercise allows us to generalize our previous results, which were  derived for equal-mass interactions  \citep{2014ApJ...784...71S, 2016arXiv160909114S, 2017arXiv170309703S}, and  robustly conclude that $>1\%$ of all BBH mergers  assembled through binary-single interactions should manifest as two-body GW captures.  As the majority of these GW captures form with high eccentricity,
our study even suggests that the binary-single interaction channel is likely to dominate the rate of high eccentricity BBH mergers
observable by LIGO.  Besides being a major motivation for developing
non-circular GW templates \citep[e.g.][]{2016PhRvD..94b4012H, 2016arXiv160905933H}, we further conclude that the notable eccentricity associated
with GW inspirals might also turn out to play a leading role in helping differentiate between the range of viable BBH merger  formation channels. The formation of electromagnetic transients through tidal two-body captures will be explored in a separate paper.

The paper is organized as follows. In Section \ref{sec:Inspirals Forming In Binary-Single Interactions}
we present a  general analytical framework for calculating the cross section and corresponding rate of two-body captures and collisions that 
form during binary-single interactions. In Section \ref{sec:Gravitational Wave Dissipation} and \ref{sec:Inspirals Compared to Post-Interaction GW Mergers}, we then
use this framework to study the absolute and relative formation rates of GW mergers. In Section \ref{sec:Formation of Binary Black Hole Mergers} we
use our findings to estimate the total number of BBH mergers assembled throughout the history of a typical GC, as well as current observable rates relevant to Advanced LIGO.
Our conclusions are then summarized in Section \ref{sec:Conclusions}.

\section{Close Encounters In Binary-Single Interactions}\label{sec:Inspirals Forming In Binary-Single Interactions}

\begin{figure}
\centering
\includegraphics[width=\columnwidth]{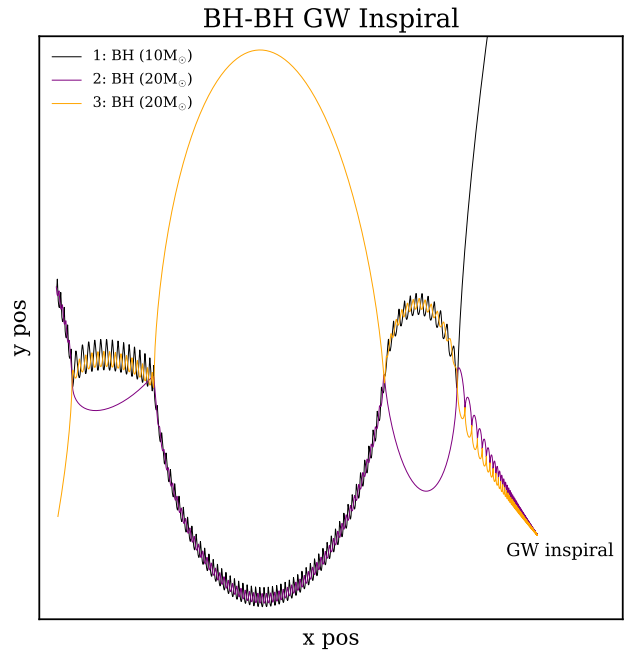}
\caption{Formation of a BBH merger through the emission of GW radiation during a resonant binary-single interaction (time increasing
from left to right). The location of the final GW inspiral is marked with `GW inspiral'. The initial SMA is set to $a_{0}=10^{-4}$ AU
for illustrative purposes. In this particular example, the two heavier BHs (purple/orange) inspiral and merge,
however, depending on the exact ICs the lighter BH (black)
could also undergo an inspiral with one of the heavier BHs. Similar inspirals can also form through tidal
interactions if one of the objects is a stellar object  \citep{2016arXiv160909114S}.
Inspirals forming in binary-single interactions generally have a very high eccentricity, which leads to particular interesting electromagnetic and GW signals.}
\label{fig:GWinsp_ill1}
\end{figure}

Binaries interacting with singles in dense stellar systems, such as GCs,
often undergo highly chaotic and resonating evolutions before reaching an end-state \citep[e.g.][]{Hut:1983js}.
During such evolutions, the three objects generally undergo several close pairwise encounters, each of which
will either result in a collision or orbital energy losses through mechanisms such as
tides and GW emission \citep[e.g.][]{2016arXiv160909114S}. While a collision simply leads
to a coalescence of the two objects \citep[e.g.][]{Fregeau:2004fj},
the fate of objects driven by orbital energy losses is less understood. However, as illustrated in \cite{2006ApJ...640..156G, 2014ApJ...784...71S, 2016arXiv160909114S},
if the energy loss is significant, the two objects will undergo a dissipative capture resulting in the formation of a compact binary. Depending on the energy loss mechanism,
this binary either ends up promptly merging or settling into a tight quasi-stable configuration.
In this paper we refer to such two-body dissipative captures as {\it inspirals}. An example of an inspiral forming through GW energy losses
is shown in Figure \ref{fig:GWinsp_ill1}. If the two objects instead undergo a passage that is smaller
than their total radii without inspiraling first, the outcome is here referred to as a \emph{collision}.

In the sections that follow we calculate the cross section and corresponding rate of inspirals and collisions forming in binary-single interactions.
We include the possibility for the three interacting objects to have different masses, which represents a major improvement to
our previous work \citep{2014ApJ...784...71S, 2016arXiv160909114S}.
This allow us to study the role of dissipative effects including tides and GW emission in binary-single systems involving realistic
combinations of white dwarfs (WDs), neutron stars (NSs), and BHs.

\subsection{Cross Sections and Formation Rates}\label{sec:Inspiral Cross Section}

We start this section  by deriving a set of quantities relevant for estimating general
cross sections and rates. Below, and in the remaining parts of the paper,
we denote an inspiral outcome between object pair $[i,j]$ by $I_{\rm  ij}$,
and a collision outcome by $C_{\rm  ij}$.

\subsubsection{Cross Section}\label{crosssection}

Imagine $N_{\rm  tot}$ uncorrelated scattering experiments
between a randomly orientated binary and an incoming single isotropically sampled across a
disk at infinity. In this case the cross section for object pair $[i,j]$ to result in a given outcome $X$, denoted by
$X_{\rm ij}$, can be estimated by the following product \citep{Hut:1983js},
\begin{equation}
\sigma_{X_{\rm ij}} = \frac{N_{X_{\rm ij}}}{N_{\rm  tot}} \times \pi b_{\rm  max}^{2},
\label{eq:sigma_i}
\end{equation}
where $N_{X_{\rm ij}}$ is the number of interactions ending as outcome $X_{\rm ij}$, and $b_{\rm  max}$ is the radius of the disk at infinity.
Here $X_{\rm ij}$ could for example denote an inspiral outcome $I_{\rm  ij}$, or a collision outcome $C_{\rm  ij}$.
The value of $b_{\rm  max}$ should be large enough for the triple system to be able to probe all
possible pathways that could result in $X_{\rm ij}$.
To determine a proper value for $b_{\rm  max}$ it is useful to work with
the corresponding pericenter distance, $r_{\rm  max}$, between the binary
center-of-mass (COM) and the incoming single. This distance relates to $b_{\rm  max}$
as follows \citep{2014ApJ...784...71S},
\begin{equation}
b_{\rm  max} = \sqrt{\frac{2Gm_{\rm  bs}r_{\rm  max}}{v_{\rm \infty}^{2}}},
\end{equation}
where we have assumed the gravitational focusing limit. Here $m_{\rm  bs}$ is the total mass of the binary-single system, and $v_{\rm \infty}$ is the initial relative
velocity between the binary and the single at infinity. The cross section $\sigma_{X_{\rm ij}}$ can now be expressed in terms of $r_{\rm  max}$ as,
\begin{equation}
\sigma_{X_{\rm ij}} = \frac{N_{X_{\rm ij}}}{N_{\rm  tot}} \times {\frac{2\pi G m_{\rm  bs}r_{\rm  max}}{v_{\rm \infty}^{2}}}.
\label{eq:sigma_i_rmax}
\end{equation}
Now, it has been shown that both inspirals and collisions predominantly form in
resonant interactions (RIs) due to their long lived chaotic
nature that makes it possible for the objects to undergo multiple close passages \citep{2014ApJ...784...71S, 2016arXiv160909114S}.
A triple system can only enter such a bound resonant state if the initial total orbital energy is negative, which is the case for
$v_{\infty} < v_{\rm c}$, where $v_{\rm c}$ is a characteristic velocity given by \citep{Hut:1983js},
\begin{equation}
v_{\rm c} = \sqrt{\frac{Gm_1m_2}{m_3(m_1+m_2)}\frac{m_{\rm bs}}{a_{0}}}.
\label{eq:v_c}
\end{equation}
Here $a_{0}$ is the semi-major axis (SMA) of the initial target binary, and the indices `1', `2', and `3' refer to the two
objects initially in the target binary and the incoming
single object, respectively (see Figure \ref{fig:ae_ill}). The limit where $v_{\infty} < v_{\rm c}$
is usually referred to as the hard-binary (HB) limit, where $v_{\infty} > v_{\rm c}$ is referred to as the soft-binary (SB) limit.
Not all initial conditions (ICs) can lead to a RI in the HB limit, and we will in this work therefore
refer to a binary-single interaction which could result in a RI 
as a \emph{close interaction} (CI). The value for $r_{\rm  max}$ which marks the limit for when interactions
with pericenter $r_{\rm  p}<r_{\rm  max}$ are all CIs is here denoted by $r_{\rm  CI}$. Although it can be shown
that no exact value of $r_{\rm  CI}$ can be defined, one can generally think of $r_{\rm  CI}$ as
representing the limit dividing democratic and hierarchical RIs \citep[e.g.][]{Hut:1993gs, 2014ApJ...784...71S}; a distance that naturally
relates to the tidal disruption distance of the binary. In this work
we take $r_{\rm  CI} = \mathscr{C}a_{0}$, where $\mathscr{C}$ is a dimensionless factor that only depends on the relative mass
ratios. With this definition of $r_{\rm  CI}$, one can now write the cross section for a CI as, 
\begin{equation}
\sigma_{\rm  CI} =  \mathscr{C} \frac{2\pi G m_{\rm  bs}a_{\rm 0}}{v_\infty^2}.
\label{eq:sigma_CI}
\end{equation}
From this  it follows that the cross section for outcome  $X_{\rm ij}$ can be written as the following product,
\begin{equation}
\sigma_{X_{\rm ij}} \approx P(X_{\rm  ij} | {\rm CI}) \times \sigma_{\rm  CI},
\label{eq:sigma_insp_1}
\end{equation}
where $P(X_{\rm  ij} | {\rm CI})$ is the probability for $X_{\rm  ij}$ to be an outcome given the
interaction is a CI. As seen, in this notation the probability $P(X_{\rm  ij} | {\rm CI})$ stands for the factor $N_{X_{\rm ij}}/N_{\rm  tot}$ in equation~\ref{eq:sigma_i}.
In the following section, we proceed to calculate the associated formation rate.

\subsubsection{Formation Rate}

The derived cross section in Section~\ref{crosssection} is especially useful for estimating the
number of outcomes $X_{\rm ij}$ forming per time interval, referred to as the rate $\Gamma_{X_{\rm ij}}$.
Following \cite{2014ApJ...784...71S}, this rate can be written for a given mass hierarchy as,
\begin{equation}
\Gamma_{X_{\rm ij}}^{(V)} \approx \frac{N_{\rm bin} N_{\rm sin}v_{\infty}}{V} \int \sigma_{X_{\rm ij}} P_{\rm bin}(a_{0}) da_{0}.
\label{eq:rate_general_eq}
\end{equation}
Here $\Gamma_{X_{\rm ij}}^{(V)}$ denotes the number of outcomes $X_{\rm ij}$ forming per unit time in a steady-state isotropic $N$-body system with volume $V$
and characteristic velocity dispersion $v_{\rm \infty}$, $N_{\rm  bin}$ and $N_{\rm  sin}$ are the total number of binaries and
singles in the system, respectively, and $P_{\rm bin}(a_{0})$ is the normalized SMA probability distribution.
We will later derive formation rates of inspirals and collisions for a few relevant distributions of $P_{\rm bin}(a_{0})$.

\subsection{Formation of Inspirals}\label{sec:Resonant Evolution of Triple Systems}

We now derive analytical expressions for the cross section and rate of inspirals
forming in binary-single interactions. For this, we first have to calculate the inspiral probability term
$P(I_{\rm  ij} | {\rm CI})$, as indicated by Equation \eqref{eq:sigma_insp_1}.
The first step in calculating $P(I_{\rm  ij} | {\rm CI})$ is to make use of
the observation that RIs generally can be described
as a series of states characterized by a binary with a bound single \citep{2014ApJ...784...71S, 2016arXiv160909114S}.
We refer to each of these states as an {\it intermediate state} (IMS),
and the binary in the state as an IMS binary. Between
each IMS the three objects undergo a strong triple interaction,
in which they semi-randomly exchange energy and orbital momentum.
As a result, a newly formed IMS binary will always have a SMA and an eccentricity
that is different from that of the initial target binary. Highly eccentric
IMS binaries can therefore form during a RI, even when the target binary is circular.
Now, if the IMS binary eccentricity is high enough, or correspondingly if the pericenter distance is small enough, the IMS
binary will be able to undergo an inspiral through the loss of orbital energy, while remaining bound to the single
in the resonance. In the next section we calculate what the initial orbital parameters of the IMS binary must be for it
to undergo such an inspiral.

\begin{figure}
\centering
\includegraphics[width=\columnwidth]{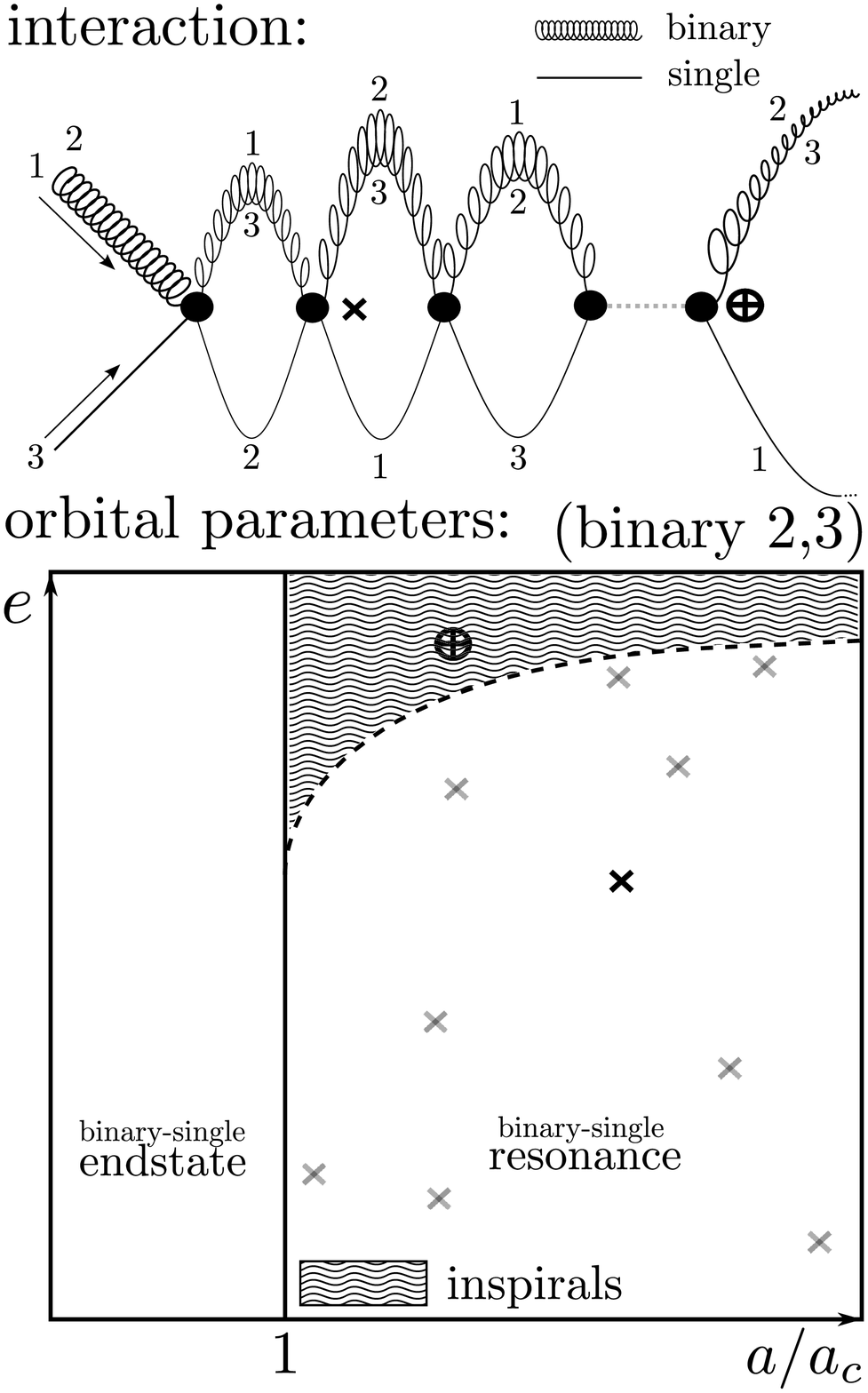}
\caption{Illustration of an inspiral forming during a resonant binary-single interaction between three
objects, where [1,2]  form the initial binary and [3] is the incoming object.
\emph{Top:} Illustration of the dynamical evolution of the three objects from initial interaction ({\it left}) to final inspiral ({\it right}).
As illustrated, a three-body RI often evolves through a series of IMSs, that are characterized by an IMS binary $[i,j]$ with a
bound single \citep{2014ApJ...784...71S, 2016arXiv160909114S}. Between
each IMS the three objects undergo a strong interaction ({\it large black dots}), where they mix and exchange energy and angular momentum.
Different binary pairs can therefore form during the interaction, each with a finite
probability for undergoing an inspiral. In this example the system evolves through various  IMS binary states until [2,3]
inspiral and merge.
\emph{Bottom:} Orbital phase space spanned by $a',e$ for IMS binary [2,3], where the wavy part illustrates the region in
which inspirals form. The black cross shows $a',e$ for the similar marked IMS binary at the top plot, where the grey crosses
illustrate a likely distribution of IMS binaries that form during the evolution marked by the small grey dots at the top plot.
The plus sign with a circle shows $a',e$ for the final [2,3] inspiral.
}
\label{fig:ae_ill}
\end{figure}

\subsubsection{The Inspiral Boundary}\label{sec: Formation of Inspirals in Orbital Phase Space}

A given IMS can be described as an IMS binary $[i,j]$ with a bound single $k$, where $\{i,j,k\}$ can be any combination of the three objects $\{1,2,3\}$. 
The IMS binary is formed with a SMA $a$ and eccentricity $e$, according to some distribution.
The first question is what combinations of $a,e$ will result in an IMS binary $[i,j]$  undergoing a successful
inspiral during the resonance while $k$ is still bound to the system.

For determining the orbital evolution of an IMS binary including orbital energy losses, we  assume a model where
the binary loses a constant amount of orbital
energy, $\Delta{E}_{\rm  p}$, during each pericenter passage \citep{2016arXiv160909114S}.
As a result, in the orbit averaged approximation, the change in orbital energy per unit time
can be written as,
\begin{equation}
\frac{dE_{\rm orb}}{dt} \approx \frac{\Delta{E}_{\rm  p}}{T_{\rm orb}} = \frac{\Delta{E}_{\rm  p}}{Gm_{\rm  ij}}\frac{\sqrt{2}}{\pi}\mu_{\rm  ij}^{-3/2}E_{\rm orb}^{3/2},
\label{eq:dEdt}
\end{equation}
where $m_{\rm  ij} = m_{\rm  i} + m_{\rm  j}$, $\mu_{\rm ij} = m_{\rm i}m_{\rm j}/m_{\rm ij}$, and $E_{\rm orb}$ ($T_{\rm orb}$) denotes the orbital energy (time) of the IMS binary.
The solution to this equation leads to a corresponding inspiral
time, $t_{\rm  insp}$, which we here define as the time it takes for the IMS binary to inspiral from its initial SMA $ = a$ to SMA $ = 0$. The solution for
$t_{\rm  insp}$ is trivially found by integration of Equation \eqref{eq:dEdt},
\begin{equation}
t_{\rm  insp} = 2\pi\sqrt{Gm_{\rm  ij}}\mu_{\rm  ij}\frac{\sqrt{a}}{\Delta{E}_{\rm  p}}.
\label{eq:t_insp_general}
\end{equation}
To keep our formalism as general as possible, we now assume that $\Delta{E}_{\rm  p}$ takes the following generic form,
\begin{equation}
\Delta{E}_{\rm  p} =  \mathscr{E}\frac{GM^2}{\mathscr{R}}\left( \frac{\mathscr{R}}{r_{\rm  p}} \right)^{\beta},
\label{eq:dE_rp_general_form}
\end{equation}
where $\mathscr{E}$ is a dimensionless normalization factor, $M$ is the characteristic mass scale, $\mathscr{R}$ is the characteristic length, and
$r_{\rm  p}$ is the IMS binary pericenter distance at the time of formation.
As will be argued, this form can be used to describe orbital energy losses from both tides and GW emission \citep[e.g.][]{2016arXiv160909114S}.
Using this generic notation one can write the inspiral time as, 
\begin{equation}
t_{\rm  insp} = 2\pi r_{\rm  p}^{\beta}\sqrt{a} \left[ \frac{{m_{\rm  ij}}\mu_{\rm  ij}\mathscr{R}^{1-\beta}}{\mathscr{E}M^{2}\sqrt{Gm_{\rm  ij}}} \right].
\label{eq:t_insp}
\end{equation}
Now, for the IMS binary to uninterrupted undergo an inspiral, its inspiral time, $t_{\rm  insp}$, must be shorter than
the time it is isolated from the bound single, $t_{\rm iso}$.
This isolation time $t_{\rm iso}$ simply equals the Keplerian orbital time of the bound single with respect to the COM of the IMS binary,
\begin{equation}
t_{\rm  iso} = 2\pi \sqrt{\frac{a_{\rm  bs}^3}{Gm_{\rm  bs}}},
\label{eq:t_iso_general}
\end{equation}
where $a_{\rm  bs}$ is the semimajor axis of the single with respect to the COM of the IMS binary.
The SMA $a_{\rm  bs}$ can be found from energy conservation by assuming the triple system did not lose orbital
energy before the formation of the IMS binary. From this assumption it follows that
\begin{equation}
\frac{m_{\rm 1}m_{\rm 2}}{2a_{\rm 0}} = \frac{m_{\rm  i}m_{\rm  j}}{2a} + \frac{m_{\rm  ij}m_{\rm  k}}{2a_{\rm  bs}},
\end{equation}
from which we find 
\begin{equation}
a_{\rm  bs} = a_{\rm 0} \left( \frac{m_{\rm  ij}m_{\rm  k}}{m_{\rm  1}m_{\rm  2}} \right) \left( \frac{a'}{a'-1} \right),
\label{eq:a_bs}
\end{equation}
where, 
\begin{equation}
a' = \frac{a}{a_{\rm c}},\;\;{\rm and, }\; a_{\rm c} = a_{\rm 0} \left(\frac{m_{\rm i}m_{\rm j}}{m_{\rm 1}m_{\rm 2}}\right).
\label{eq:diff_a'}
\end{equation}
Inserting the relation for $a_{\rm  bs}$ into Equation \eqref{eq:t_iso_general}, the isolation time can now be expressed as,
\begin{equation}
t_{\rm iso} = 2\pi \frac{a_{\rm 0}^{3/2}}{\sqrt{Gm_{\rm bs}}}\left( \frac{m_{\rm ij}m_{\rm k}}{m_{\rm 1}m_{\rm 2}} \right)^{3/2} \left( \frac{a'}{a'-1} \right)^{3/2}.
\label{eq:t_iso}
\end{equation}
By equating $t_{\rm insp}$ and $t_{\rm iso}$ given by Equation \eqref{eq:t_insp} and \eqref{eq:t_iso}, respectively,
we can then find the following conditional relation between the IMS binary orbital parameters $a,e$,
\begin{equation}
\epsilon_{I_{\rm ij}} = \mathscr{E}^{1/\beta} \mathscr{M} \left({a_{\rm 0}}/{\mathscr{R}}\right)^{(1/\beta-1)} \mathscr{G}(a',\beta),
\label{eq:e_insp}
\end{equation}
where
\begin{equation}
\mathscr{G}(a',\beta) =  a'^{(1/\beta-1)}\left(a'-1\right)^{-3/(2\beta)},
\end{equation}
\begin{equation}
\epsilon_{I_{\rm ij}} = (1-e_{I_{\rm ij}}),
\end{equation}
$e_{I_{\rm ij}}$ is the eccentricity for which $t_{\rm insp} = t_{\rm iso}$ for a
given $a'$, and $\mathscr{M}$ is a dimensionless term given by,
\begin{equation*}
\mathscr{M} = \left(\frac{m_{\rm 1}m_{\rm 2}}{m_{\rm i}m_{\rm j}}\right) \left[ \left(\frac{M}{m_{\rm bs}}\right)^{2} \left(\frac{m_{\rm bs}}{\mu_{\rm ij}}\right)^{{3}/{2}} \left(\frac{m_{\rm k}m_{\rm k}}{m_{\rm 1}m_{\rm 2}}\right) \left(\frac{m_{\rm ij}}{m_{\rm k}}\right)^{{1}/{2}} \right]^{1/\beta}.
\label{eq:M_mass_term}
\end{equation*}

The derived relation between $e_{I_{\rm ij}}$ and $a'$ given by Equation \eqref{eq:e_insp}
defines a boundary in orbital phase space which encloses the region in which inspirals can form, as illustrated  in Figure \ref{fig:ae_ill}.
This inspiral region, which  for the IMS binary in question we denote $\mathcal{R}_{\rm I}$, is bounded along the $a'$-axis
by a lower limit $a'_{\rm l}$, and an upper limit $a'_{\rm u}$. The
value\footnote{The lower limit should in fact
be marginally higher than $1$, since $a'_{\rm l}=1$ corresponds to an infinite interaction time.
However, uncertainties in how to define the upper limit makes this correction irrelevant for the scales we consider in this work.} of $a'_{\rm l}$ is $\approx 1$,
where the value for $a'_{\rm u}$ relates to the limit for when the resonant triple system no longer can be
described as an IMS. That is, when $a_{\rm bs} \approx a_{0}$.
One way of estimating $a'_{\rm u}$ is by considering the force ratio $f_{\rm tid} = F_{\rm tid}/F_{\rm bin}$,
where $F_{\rm tid}$ is the tidal force the bound single exerts on the IMS binary, and $F_{\rm bin}$ is the IMS binary's binding force \citep{Fregeau:2004fj},
\begin{align}
F_{\rm tid}  & \approx \frac{1}{2}\frac{Gm_{\rm ij}m_{\rm k}}{a_{\rm bs}^{2}} \frac{a}{a_{\rm bs}},\\
F_{\rm bin} & \approx \frac{1}{4}\frac{Gm_{\rm i}m_{\rm j}}{a^{2}}.
\end{align}
By the use of Equations \eqref{eq:a_bs} and \eqref{eq:diff_a'}, one can solve for $a'$ as a function of the
mass hierarchy and the value of $f_{\rm tid}$. In this work we take $a'_{\rm u}$ to be the value of $a'$ for which\footnote{Our results depend only weakly
on the precise value as $a'_{\rm u}-1 \propto f_{\rm tid}^{1/3}$.}
 $f_{\rm tid} = 0.5$. To summarize, the two boundaries $a'_{\rm l}$ and $a'_{\rm u}$ approximately take the values,
\begin{align}
a'_{\rm l}  & \approx 1,\\
a'_{\rm u} & \approx 1 + \left( \frac{1}{2}\frac{m_{\rm k}}{\mu_{\rm ij}} \right)^{2/3}.
\end{align}
In the following section we describe how to relate the inspiral region $\mathcal{R}_{\rm I}$ derived  here to the probability
term $P(I_{\rm ij} | {\rm CI})$.

\subsubsection{Inspiral Probability}\label{sec:Inspiral Probability}

The probability $P(I_{\rm ij} | {\rm CI})$ can be written as 
\begin{equation}
P(I_{\rm ij} | {\rm CI}) \approx \langle N_{\rm IMS} \rangle P(\mathcal{R}_{\rm I}),
\end{equation}
where $\langle N_{\rm IMS} \rangle$ denotes the average number of IMS binaries $[i,j]$ formed in a CI,
and $P(\mathcal{R}_{\rm I})$ is the probability for a single IMS binary $[i,j]$ to be formed with [$a',e$] within the inspiral region $\mathcal{R}_{\rm I}$.
We have here  assumed that $P(\mathcal{R}_{\rm I}) \ll 1$, and that the sampling
of IMS binaries is not correlated.

\begin{figure}
\centering
\includegraphics[width=\columnwidth]{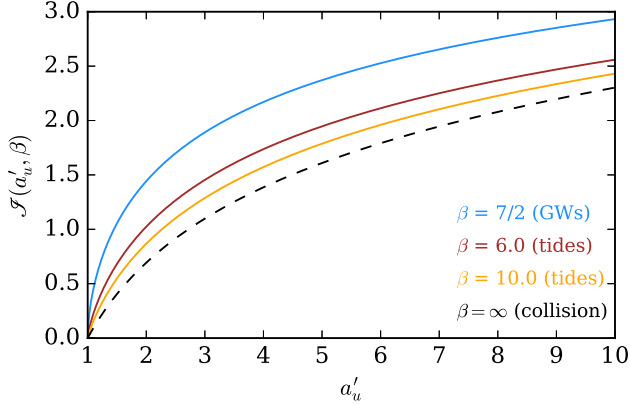}
\caption{Value of the inspiral integral $\mathscr{I}(a'_{\rm u}, \beta)$ derived by numerical integration of Equation \eqref{eq:I_apu}, as a
function of $a'_{\rm u}$ for a few relevant values of $\beta$. The values $\beta = 7/2,\ 6,\ 10, \infty$,
relate to the description of GW emission, tides ($n=1.5$ polytrope), tides ($n=3.0$ polytrope), and
collisions, respectively \citep[see e.g.][]{2016arXiv160909114S}. For this plot we have set $a'_{\rm l}=1$.} 
\label{fig:insp_integral}
\end{figure}

To proceed we assume that the [$a',e$] distribution is uniform at high eccentricity.
This essential assumption was first shown to hold for the equal mass case by \cite{2014ApJ...784...71S}, however,
as we will show, we find that the distribution also seems to be approximately uniform for the unequal mass
case. Figure \ref{fig:IMS_insp_ill_SIM} shows an example of this.
From this assumption it follows that the probability $P(\mathcal{R}_{\rm I})$ can be factorized as
\begin{equation}
P(\mathcal{R}_{\rm I})  \approx  P(\mathcal{R}_{\rm U}) P(\mathcal{R}_{\rm I} | \mathcal{R}_{\rm U}),
\end{equation}
where $\mathcal{R}_{\rm U}$ denotes the region where the [$a',e$] sampling is assumed uniform,
$P(\mathcal{R}_{\rm U})$ is the probability for the IMS binary to form
inside $\mathcal{R}_{\rm U}$, and $P(\mathcal{R}_{\rm I} | \mathcal{R}_{\rm U})$ is the probability for the
IMS binary to form inside $\mathcal{R}_{\rm I}$ given that it was  formed within  $\mathcal{R}_{\rm U}$.
This implies that the probability $P(I_{\rm ij} | {\rm CI})$ can be written as
\begin{equation}
P(I_{\rm ij} | {\rm CI}) \approx  {A_{\rm I}} \left[ \frac{\langle N_{\rm IMS} \rangle {P(\mathcal{R}_{\rm U})}}{{A_{\rm U}}} \right],
\label{eq:PI_CI_NA}
\end{equation}
where $A_{\rm I}$ and $A_{\rm U}$ are the areas of the $\mathcal{R}_{\rm I}$ and $\mathcal{R}_{\rm U}$ regions, respectively.
In what follows we define 
\begin{equation}
\mathscr{N}\equiv \left[ \frac{\langle N_{\rm IMS} \rangle {P(\mathcal{R}_{\rm U})}}{{A_{\rm U}}} \right]
\end{equation}
and  note that the value of $\mathscr{N}$ do not depend to leading order on
any finite size effects, orbital energy losses, the absolute mass scale or the initial SMA $a_{\rm 0}$, as further discussed
in Section \ref{sec:Summary On Our Analytical Model}. Any effects related to orbital energy losses
are therefore fully encoded in the area factor $A_{\rm I}$. This factor is given by the integral
of $\epsilon_{I_{\rm ij}}$ over $a'$, from $a'_{\rm l}$ to $a'_{\rm u}$,
\begin{equation}
A_{\rm I} \approx \int_{a'_{\rm l}}^{a'_{\rm u}} \epsilon_{I_{\rm ij}} da' = \mathscr{E}^{1/\beta} \mathscr{I} \mathscr{M} \left({a_{\rm 0}}/{\mathscr{R}}\right)^{(1/\beta-1)},
\label{eq:AIij}
\end{equation}
where $\mathscr{I}$ denotes the value of the integral,
\begin{equation}
\mathscr{I} = \int_{a'_{\rm l}}^{a'_{\rm u}} \mathscr{G}(a',\beta) da'.
\label{eq:I_apu}
\end{equation}
This integral has  no analytical solution and must  be evaluated numerically.
We nonetheless  find that the solution can be accurately approximated by
the following functional form
\begin{equation}
\mathscr{I} \approx \frac{1.05}{(1-1.7/\beta)}  \frac{\text{ln}{(a'_{\rm u})}}{(a'_{\rm u}-1)^{1/(\beta + 1)}}.
\end{equation}
A few examples of $\mathscr{I}$ as a function of $a'_{\rm u}$ and $\beta$ are shown in Figure \ref{fig:insp_integral}.
By substituting the expression for $A_{\rm I}$, given by Equation \eqref{eq:AIij}, into the relation for $P(I_{\rm ij} | {\rm CI})$,
given by Equation \eqref{eq:PI_CI_NA}, we find
\begin{equation}
P(I_{\rm ij} | {\rm CI}) \approx \mathscr{N} \mathscr{E}^{1/\beta} \mathscr{I} \mathscr{M} \left({a_{\rm 0}}/{\mathscr{R}}\right)^{(1/\beta-1)}.
\label{eq:PIij_CI_final}
\end{equation}
We can now make use of this relation to  estimate  the inspiral cross section and corresponding
formation rate.

\subsubsection{Inspiral Cross Section}\label{sec:The Inspiral Cross Section}

The cross section for object pair [$i,j$] to undergo an inspiral, here referred to as the inspiral cross section
$\sigma_{I_{\rm ij}}$, is given by the product of the inspiral probability
and the CI cross section, $P(I_{\rm ij} | {\rm CI}) \times \sigma_{\rm CI}$, as denoted in Equation \eqref{eq:sigma_insp_1}.
Making use  of Equation \eqref{eq:PIij_CI_final} and Equation \eqref{eq:sigma_CI}, we can then find
\begin{equation}
\sigma_{I_{\rm ij}} \approx \sigma_{\rm \mathscr{R}_{\rm ij}} \times \left[\mathscr{E}^{1/\beta} {\mathscr{I}}' \mathscr{M}' \left({a_{\rm 0}}/{\mathscr{R}}\right)^{1/\beta}\right],
\label{eq:final_sigma_Iij}
\end{equation}
where we have introduced the cross section term $\sigma_{\rm \mathscr{R}_{\rm ij}}$, and have defined the following quantities
\begin{align}
\mathscr{M}' & = \mathscr{M} \left(\frac{m_{\rm i}m_{\rm j}}{m_{\rm 1}m_{\rm 2}}\right),\\
\mathscr{I}'   & = \mathscr{I} \frac{1}{\ln(a'_{\rm u})}.
\end{align}
The term $\sigma_{\rm \mathscr{R}_{ij}}$ equals the cross section for object pair [$i,j$] to pass each other
within a fixed distance $\mathscr{R}$ in the limit where no energy loss terms are included in the EOM.
As later described in Section \ref{sec:Collision Cross Section}, the value for $\sigma_{\rm \mathscr{R}_{ij}}$
can be derived using Equation \eqref{eq:cs_Rij}, and is therefore proportional to the collision cross section.
As a result, the terms in the square parenthesis in Equation \eqref{eq:final_sigma_Iij}
represent the leading order effect from introducing energy loss terms into the finite-size, unequal-mass binary-single problem.

\subsubsection{Inspiral Rate}\label{sec:Inspiral Rate}

Having derived an analytical solution to the inspiral cross section, we can then proceed to 
calculate expressions for the corresponding formation rates. Here we work out rates for two representative
cases.

{\it Field Interactions.}
Binaries in the field are often found to be uniformly distributed in $\log(a_{0})$ \citep[e.g.][]{2004ApJ...601..289C, 2007AJ....133..889L},
a result also known as Oepiks law. For target binaries following such a distribution between SMA limits $a_{\rm  min}$ and $a_{\rm  max}$, the
resultant rate of inspirals can  be found by direct integration of Equation \eqref{eq:rate_general_eq} using the cross section
solution derived in Equation \eqref{eq:final_sigma_Iij}, 
\begin{equation}
\Gamma_{I_{\rm ij}}^{(V)} \approx \frac{N_{\rm  bin}N_{\rm  sin}v_{\rm \infty}}{V} \times \frac{\sigma_{I_{\rm ij}}(a_{\rm  max}) - \sigma_{I_{\rm ij}}(a_{\rm  min})}{\ln(a^{1/\beta }_{\rm  max}/a^{1/\beta }_{\rm  min})}.
\label{eq:Gamma_insp_general}
\end{equation}
In systems where the SMA distribution is unknown, one often assumes this distribution for simplicity.

{\it Cluster Interactions.}
Binaries in dynamical systems, such as GCs, have at late times a SMA distribution which  can be 
described by a Gaussian function  in $\log(a_{0})$, as recently pointed out by \cite{2016PhRvD..93h4029R}.
In this case, the inspiral rate can be written as
\begin{equation}
\Gamma_{I_{\rm ij}}^{(V)} \approx \frac{N_{\rm  bin}N_{\rm  sin}v_{\rm \infty}}{V} \times \sigma_{I_{\rm ij}}(a_{\rm  c})\exp\left( \frac{\ln(10)^{2}}{2}\frac{{s^2}}{\beta^2}\right),
\label{eq:Gamma_I_gaussian}
\end{equation}
where $a_{\rm c}$ is the SMA of the Gaussian peak, and $s$ is the standard deviation.
As can be seen, $s$ must be larger than $\beta$ for the assumption of a Gaussian distribution to modify the simple estimate that assumes that all target binaries 
have the same initial SMA $a_{0} = a_{\rm c}$. As $\log(\sigma_{\rm I}) \propto (1/\beta)\log(a_{0})$, the derived rates are only sensitive to the shape of the Gaussian distribution 
when the inspiral cross section changes significantly over  its intrinsic width.
We note that since $\beta > 3.5$ for both tides and GW emission \citep{2016arXiv160909114S}, this condition is in fact rarely met in
a typical GC system where $s \approx 0.5$, as inferred from the distributions shown in \cite{2016PhRvD..93h4029R}.

\subsection{Collisions}\label{sec:Collision Cross Section}

We now turn our attention  to collisions.
A newly formed IMS binary [$i,j$] will undergo a collision if its pericenter distance $r_{\rm p}$ is smaller
than a characteristic collisional distance $R_{\rm C}$.
The maximum pericenter distance from which a collision can happen is therefore per construction  given by
$r_{\rm p}={R_{\rm C}}$. This limit is met when the orbital parameters of the IMS binary are such that,
\begin{equation}
{R_{\rm C}} = a(1-e),
\label{eq:Ra_1e}
\end{equation}
from which it follows that 
\begin{equation}
\epsilon_{C_{\rm ij}} = \left( \frac{m_{\rm 1}m_{\rm 2}}{m_{\rm i}m_{\rm j}} \right) \frac{{R}_{\rm C}}{a_{\rm 0}} \frac{1}{a'},
\label{eq:eps_R}
\end{equation}
where 
\begin{equation}
\epsilon_{C_{\rm ij}} = (1-e_{C_{\rm ij}}),
\end{equation}
and $e_{C_{\rm ij}}$ is the eccentricity that satisfies Equation \eqref{eq:eps_R} for a given $a'$.
The relation between $\epsilon_{C_{\rm ij}}$ and $a'$ defines a boundary in orbital phase space,
which encloses the region in which IMS binaries will undergo a collision ($r_{\rm p}<{R_{\rm C}}$). The area of this region, denoted $A_{\rm C}$, is given by
\begin{equation}
A_{\rm C} \approx \int_{a'_{\rm l}}^{a'_{\rm u}} \epsilon_{C_{\rm ij}} da' = \frac{{R}_{\rm C}}{a_{\rm 0}} \left( \frac{m_{\rm 1}m_{\rm 2}}{m_{\rm i}m_{\rm j}} \right) \ln(a'_{\rm u}).
\end{equation}
Following the same procedure as we did for  inspirals, the probability for a collision outcome
${C_{\rm ij}}$ provided that  the interaction is a CI can now be factorized as $P({C}_{\rm ij} | {\rm CI}) \approx A_{\rm C}\mathscr{N}$. It then
follows that the collision cross section can be written as
\begin{equation}
\sigma_{{C}_{\rm ij}} \approx \mathscr{F} \left(\frac{2 \pi G m_{\rm bs}{R}_{\rm C}}{v_{\rm \infty}^{2}}\right) \left(\frac{m_{\rm 1}m_{\rm 2}}{m_{\rm i}m_{\rm j}}\right) \ln(a'_{\rm u}),
\label{eq:cs_Rij}
\end{equation}
where we have defined $\mathscr{F} \equiv \mathscr{C} \times \mathscr{N}$, a factor that will be discussed in more detail in Section \ref{sec:Summary On Our Analytical Model}.
The above equation for $\sigma_{{C}_{\rm ij}}$ is similar to Equation \eqref{eq:final_sigma_Iij} but evaluated at
$R_{\rm C} = \mathscr{R}$. In fact, our derived form for $\sigma_{{C}_{\rm ij}}$ can be used to estimate the
cross section for any fixed distance close encounter, such as, e.g., a tidal disruption event. It is here worth noting that
the collision cross section is independent of the initial SMA $a_{0}$, and that the inspiral cross section approaches the collision
cross section for $\beta \rightarrow \infty$.

\subsubsection{Collision Rate}

The rate of collisions can be estimated by integrating Equation \eqref{eq:rate_general_eq}
using the collision cross section  in Equation \eqref{eq:cs_Rij}. However, in this case, the solution
is particularly simple as the collision cross section does not depend on $a_{0}$. This implies that the corresponding collision rate $\Gamma_{C_{\rm ij}}$ is independent
of the SMA distribution. As a result, the rate of collisions takes the following simple form
\begin{equation}
\Gamma_{C_{\rm ij}}^{(V)} \approx \frac{N_{\rm  bin}N_{\rm  sin}v_{\rm \infty}}{V_{\rm }} \times \sigma_{C_{\rm ij}}.
\label{eq:Gamma_coll_general}
\end{equation}

\subsection{Inspirals Relative to Collisions}\label{sec:Inspirals Relative to Collisions}

With our derived inspiral and collision cross sections, we are now in a position to understand their relative formation rates. Assuming the inspiraling binaries also merge,
it is of particular interest to explore whether or not  inspirals are expected to 
significantly contribute  to the merger rate.
We do so by  considering the cross section ratio between inspirals and  collisions
\begin{equation}
\frac{\sigma_{I_{\rm ij}}}{\sigma_{C_{\rm ij}}} \approx \left( \frac{\mathscr{R}}{{R_{\rm C}}} \right) \times \left[{\mathscr{E}^{1/\beta}} {\mathscr{I}'} {\mathscr{M}}'
\left({{a_{\rm 0}}}/{{\mathscr{R}}}\right)^{1/\beta} \right].
\label{eq:relrate_merg_colldistrup}
\end{equation}
We first note that the factor $\mathscr{F}$ cancels out, which allows us to derive an analytical solution. This is especially important for studying the role of tides,
as the tidal inspiral cross section and the collision cross section are often comparable \citep{2016arXiv160909114S}.

The above expression shows that the number of inspirals
generally increases relative to that of collisions with increasing  $a_{\rm 0}$  and decreasing $R_{\rm C}$.
The effect of orbital energy losses is thus  largest in interactions
involving wide binaries and dense objects such as WDs, NSs, and BHs. A similar conclusion was reached
for the equal mass case in \cite{2014ApJ...784...71S, 2016arXiv160909114S}.
This counter intuitive scaling also explains why earlier studies on, e.g., tidal interactions not have
been able to resolve these notable differences between collisions and inspirals \citep[see e.g.][]{2010MNRAS.402..105G}.
The relevance and applicability of the general analytical  formalism derived here  is discussed below.

\subsection{The Validity of Our Analytical Formalism}\label{sec:Summary On Our Analytical Model}
Due to the highly chaotic nature of the three-body problem in the HB limit, a simple analytical description,
similar to the one that exists in the SB limit \citep{Hut:1983by}, is not easily derived.
Here we have outlined  a simple approach that allows one to calculate the cross section
for  a variety of  outcomes that arise  when finite size effects and orbital energy loss term corrections are taken into account.
Our  solutions make use of  the fact  that the [$a,e$] distribution is
uniform at high eccentricity, which has been validated with the use of numerical experiments \citep{2016arXiv160909114S}.

We also note that the scalings in  Equation \eqref{eq:final_sigma_Iij} can be derived, as argued in \cite{2016arXiv160909114S}, by considering the physical attributes  of the interaction.  
For two objects to undergo an inspiral, the orbital energy loss during the first passage, $\Delta{E_{\rm p}}$, must be a notable fraction of the initial orbital energy, $E_{0}$.
By equating $\Delta{E_{\rm p}}$ and $E_{0}$ and solving for the corresponding pericenter distance, which we denoted here as $r_{\rm cap}$, one finds
\begin{equation}
r_{\rm cap} \propto \mathscr{R}\left(a_{0}/\mathscr{R}\right)^{1/\beta}.
\label{eq:rcap}
\end{equation}
If the inspiral leads to a merger, one can  think of $r_{\rm cap}$ as an effective stellar size. As seen in Equation~\eqref{eq:rcap},  $r_{\rm cap}$ is not just proportional to the size of the object,
but scales with $(a_{0}/\mathscr{R})^{1/\beta}$. This scaling is  similar to the one  found in Equation \eqref{eq:final_sigma_Iij}, giving further credence to our analytical formalism.

We note that we are yet  to calculate a precise expression for $\mathscr{F}$, which 
relates to the difficulties associated  with calculating the full IMS binary [$a',e$] distribution. However, due to the classical scale-free
nature of the three-body problem \citep{Hut:1983js}, one can show that $\mathscr{F}$ depends only on the relative mass ratio and not on 
the absolute mass scale, initial SMA, finite sizes or the precise orbital energy loss mechanism.
The results reported in this paper have all been calibrated using numerical experiments, from which
we  have found, for example,  that $\mathscr{F} \approx 6.0$ for the equal mass case. When comparing
the relative rate of  inspirals to collisions  $\mathscr{F}$  cancels out, which allows  for a full analytical estimate to be derived.  

In general we find that the set of scalings  and analytical results derived here agree very well with results
from full $N$-body simulations \citep[e.g.][]{2014ApJ...784...71S, 2016arXiv160909114S}.
In the following sections we will use our analytical model to gain further insight into the dynamical and observational importance of GW inspirals.

\section{Gravitational Wave Inspirals}\label{sec:Gravitational Wave Dissipation}

When GR effects are included in the binary-single problem, a close passage between any two of the three objects
will lead to GW emission. In this process orbital energy
is carried out of the system, and can as a result lead to an inspiral and subsequent merger between the
two objects \citep{2006ApJ...640..156G, 2014ApJ...784...71S, 2016arXiv160909114S}.
In the remaining parts of this paper we will discuss several aspects related to these GW mergers
that we generally refer to as {\it GW inspirals}. For this, we will make use of our analytical
framework from Section \ref{sec:Inspirals Forming In Binary-Single Interactions},
as well as full numerical $N$-body simulations including PN correction terms.

From an astrophysical perspective, GW inspirals are extremely exciting,
as they often enter the LIGO band with non-zero eccentricity \citep{2006ApJ...640..156G, 2014ApJ...784...71S}.
In fact, recent studies indicate that they might even dominate the population of observable high
eccentricity BBH mergers \citep{2017arXiv170309703S}.
Together with the spin distribution \citep[e.g.][]{2016ApJ...832L...2R, 2016MNRAS.462..844K, 2017arXiv170200885Z}, the eccentricity distribution
is expected to play a key role in constraining the origin of BBH mergers
observable by LIGO and next generation GW observatories \citep[e.g.][]{2017arXiv170208479C}. High eccentricity NS-NS mergers are also promising probes
of the NS equation of state \citep[e.g.][]{2012ApJ...760L...4E, 2014ARA&A..52..661L, 2015ApJ...807L...3E, 2015PhRvD..92l1502P, 2016arXiv160900725E, 2016PhRvD..93b4011E}.

To provide concrete predictions and scaling solutions relevant for LIGO, we 
first  focus here on the case where all three interacting objects are equal mass BHs.
Although this represents an idealized scenario, recent numerical studies do in fact indicate that a
significant fraction of the observable population of BBH mergers forming in GCs likely
form dynamically through similar mass binary-single interactions \citep{2016PhRvD..93h4029R, 2017arXiv170309703S}.
The explanation for this is that mass segregation causes heavier BHs to reach the GC center faster than lighter ones,
BHs of similar mass are thus more likely to be found in the GC center at similar ages in the evolution of the cluster \citep[e.g.][]{2016PhRvD..93h4029R, 2017arXiv170301568P}.
Subsequent binary-single interactions also tend to keep BHs of similar mass together \citep{Sigurdsson:1993jz}.

In the sections that follow we start by introducing the GW energy loss term. We then 
explore how frequent GW inspirals are compared to collisions in encounters involving
compact objects. After this we derive a set of solutions to the equal mass case. This
is followed by a study on how slight changes to the mass hierarchy affect the GW inspiral cross section.
In the last section we use our solutions to estimate the present day GW inspiral rate observable by LIGO.

\subsection{GW Energy Loss Term}\label{sec:GW Energy Loss Term}

The energy emitted through GW radiation during a single pericenter passage between object pair $[i,j]$ is at quadrupole order
in the high eccentricity limit given by \citep{Peters:1964bc, Hansen:1972il},
\begin{equation}
\Delta{E}_{\rm  GW} \approx \frac{85\pi}{12\sqrt{2}}\frac{G^{7/2}}{c^{5}}\frac{m_{\rm i}^{2}m_{\rm j}^{2}m_{\rm ij}^{1/2}}{r_{\rm p}^{7/2}}.
\label{eq:deltaE_GR}
\end{equation}
In this limit, one finds that
$\Delta{E}_{\rm  GW}$ can also be expressed by the general form given by Equation \eqref{eq:dE_rp_general_form}
by setting,
\begin{equation}
\mathscr{E} = \frac{85\pi}{96},\  M=\mu_{\rm ij},\  \mathscr{R}=\frac{2Gm_{\rm ij}}{c^{2}},\  \beta = \frac{7}{2}.
\label{eq:E_M_R_beta_GWs}
\end{equation}
With these substitutions the derived relations from Section \ref{sec:Inspirals Forming In Binary-Single Interactions}
can now be used to study GW inspirals. Below we proceed by stuyding how frequent GW inspirals
are compared to standard collisions.

\subsection{GW Inspirals Relative to BH and NS Collisions}\label{sec:GW Inspirals Relative to BH and NS Collisions}

The final outcome of both a GW inspiral and a collision is a merger. In this
section we explore which of the two channels is expected to dominate the total merger rate.
As an example, in main sequence star interactions the collision cross section
is typically higher than the tidal inspiral cross section, where for interactions involving high mass WDs the two cross sections are of
similar order \citep{2016arXiv160909114S}.
Here we are interested  in calculating  the ratio of GW inspirals to collisions  for interactions involving BHs and NSs, and how sensitively it
depends on the initial SMA, $a_{0}$, and the mass hierarchy. For this study, we use our analytical estimate for
${\sigma_{I_{\rm ij}}}/{\sigma_{C_{\rm ij}}}$ given by Equation \eqref{eq:relrate_merg_colldistrup}. Two examples are worked out below.

\begin{figure}
\centering
\includegraphics[width=\columnwidth]{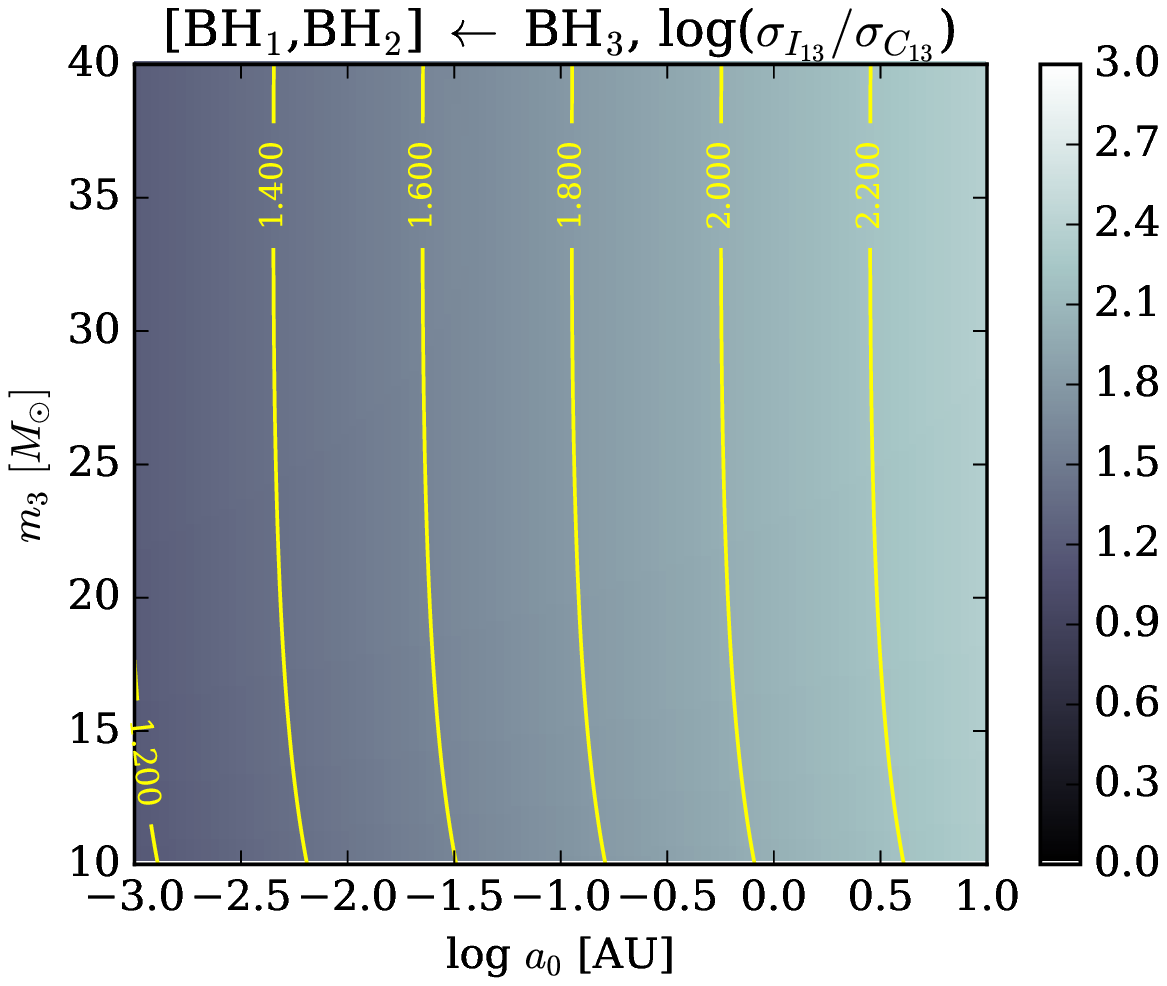}
\includegraphics[width=\columnwidth]{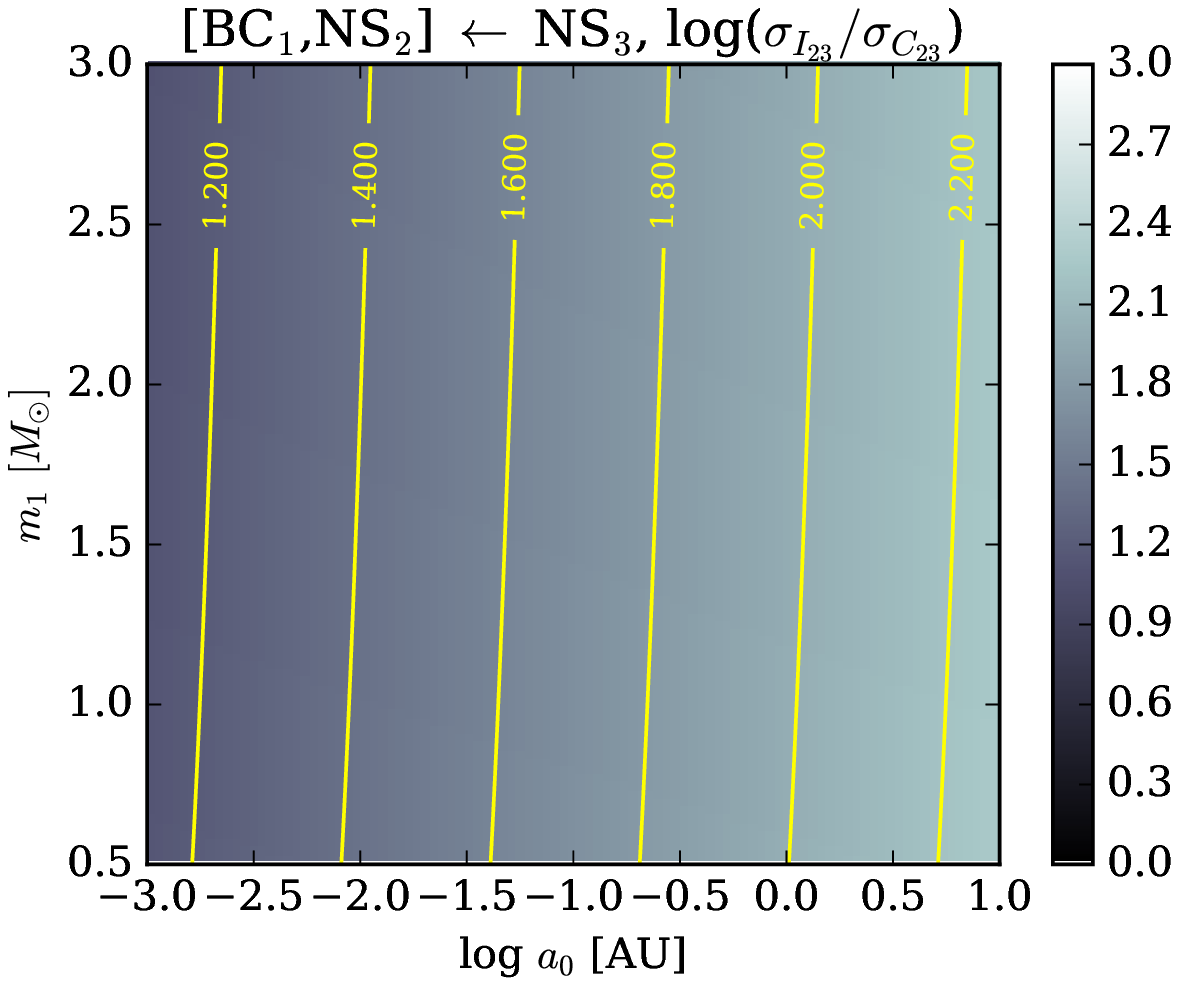}
\caption{The contours show the log of the ratio between the GW inspiral and the collision cross
sections, $\text{log}({\sigma_{I_{ij}}}/{\sigma_{{C}_{ij}}})$, for
the two interaction channels discussed in Section \ref{sec:GW Inspirals Relative to BH and NS Collisions}.
The top plot is for [BH$_{1}$, BH$_{3}$] mergers (Section \ref{sec:BHBH_mergers}), where the bottom plot shows
results for [NS$_{2}$, NS$_{3}$] mergers (Section \ref{sec:NSNS_mergers}). As seen,
in both cases the cross section of GW inspirals is about $100$ times larger than the classical collision cross section.
This illustrates that for compact objects, the GW inspiral channel completely dominates the formation rate of mergers forming in
binary-single interactions.}
\label{fig:sigmaI_sigmaR_BHNS}
\end{figure}

\subsubsection{BH-BH Mergers}\label{sec:BHBH_mergers}
We start by considering the following triple BH binary-single interaction,
\begin{equation}
[\text{BH}_{1}, \text{BH}_{2}] \leftarrow \text{BH}_{3},
\end{equation}
where the corresponding masses are $m_{1} = 30M_{\odot}$, $m_{2} = 20M_{\odot}$, and $m_{3}$ is
varied between $10M_{\odot} - 40M_{\odot}$.
The $\log$ of the cross section ratio of GW inspirals to collisions for object pair $[1,3]$, $\log({\sigma_{I_{13}}}/{\sigma_{C_{13}}})$,
as a function of the initial SMA $a_{0}$ and $m_{3}$, is shown in the {\it top panel} of Figure \ref{fig:sigmaI_sigmaR_BHNS}.
As can be seen, the GW inspiral cross section is, for this example, $\approx 10^{2}$ times higher than the collision cross section.

\subsubsection{NS-NS Mergers}\label{sec:NSNS_mergers}
We now consider a binary-single interaction involving two NSs, and an unspecified binary companion (BC),
\begin{equation}
[\text{BC}_{1},\text{NS}_{2}] \leftarrow \text{NS}_{3},
\end{equation}
where the corresponding masses are $m_{2} = 1.4M_{\odot}$,
$m_{3} = 1.4M_{\odot}$, and $m_{1}$ is varied between $0.5M_{\odot} - 3M_{\odot}$. Our chosen range of $m_{1}$ covers a few astrophysical
interesting configurations, from BC$_{1}$ representing a WD (0.5$M_{\odot}$) to a heavy NS (3$M_{\odot}$).
The log of the cross section ratio of [NS, NS] GW inspirals to collisions, $\log({\sigma_{I_{23}}}/{\sigma_{C_{23}}})$,
as a function of the initial SMA $a_{0}$ and $m_{1}$, is shown in the {\it bottom panel} of Figure \ref{fig:sigmaI_sigmaR_BHNS}. In this case,
even  for a NS, which has a slightly larger physical radius than a BH, GW inspirals clearly dominate over classical collisions.

\subsubsection{GW Inspirals versus Collisions for Compact Objects}

The number of compact object mergers that dynamically form during resonant three-body interactions is completely
dominated by GW inspirals, and not by classical collisions. We find this to be true for both BH and NS mergers.
This result is not that surprising when considering Equation \eqref{eq:relrate_merg_colldistrup}, which says that the cross section
ratio approximately equals the initial SMA $a_{0}$ over the Schwarzschild radius of the two merging BHs to the power of $2/7$. This ratio
is a large number for typical astrophysical systems, where $a_{0}$ is about 1 AU ($\approx 10^{8}$ km) and the BH Schwarzschild radius is
$\approx 10^{2}$ km. It is thus crucial to have PN terms included  in the $N$-body code for estimating a meaningful rate of
compact object mergers forming during three-body interactions.
As an example, \cite{2016ApJ...816...65A} did not have
PN terms in their few-body code, which likely led to an underestimation of their {\it in-cluster} merger rate by a
factor of $\approx 10^{1}-10^{2}$ (see Table 1 in \cite{2016ApJ...816...65A}). Because of this, their rate reported
for high eccentricity BBH mergers forming through binary-single interactions is far too low, as we argued in \citet{2017arXiv170309703S}. 

For this reason, in the rest of the paper, we will mainly focus on GW inspirals without
making any further inferences  for collisions. In fact, the collision population is included
in our analytical estimation of GW inspirals, as the inspiral area $A_{\rm I}$ overlaps with the collision area $A_{\rm C}$
in orbital phase space \citep[e.g.][]{2016arXiv160909114S}.

\subsection{Equal Mass Interactions}\label{sec:GW Inspirals in the Equal Mass Limit}

In this section we calculate the GW inspiral cross section assuming all three objects have the same mass $m$.
As argued earlier, this is an idealized case, but is still likely to provide a reasonable description of
the dynamics leading to the majority of BBH mergers forming in GCs observable by LIGO.
We note that the equal mass case was also studied in \cite{2014ApJ...784...71S, 2016arXiv160909114S}, we therefore keep the following section
concise. The results presented here  will be used in later sections for estimating absolute and relative BBH merger rates.

The inspiral boundary and corresponding inspiral cross section
can be calculated using Equations \eqref{eq:e_insp} and \eqref{eq:final_sigma_Iij}, respectively.
By  further making use of Equation \eqref{eq:E_M_R_beta_GWs},
we find  that the GW inspiral boundary can be written as,
\begin{equation}
{\epsilon}_{I_{\rm ij}} \approx  C_{\rm  GW} \times \frac{G^{{5}/{7}}m^{{5}/{7}}}{c^{10/7}a_{\rm 0}^{{5}/{7}}} \mathscr{G}(a',\beta = 7/2),
\label{eq:eps_GW_EM}
\end{equation}
and the corresponding GW inspiral cross section as,
\begin{equation}
{\sigma}_{I_{\rm ij}} \approx C_{\rm  GW} \times 6\pi{\mathscr{I}} {\mathscr{F}} \frac{G^{{12}/{7}} m^{12/7}a_{\rm 0}^{2/7}}{c^{10/7}v_{\rm \infty}^{2}},
\label{eq:cs_GWinsp_EM}
\end{equation}
where $C_{\rm  GW}$ is a constant given by, 
\begin{equation}
C_{\rm  GW} = \left(\frac{85\pi}{3\sqrt{3}}\right)^{2/7}.
\end{equation}
The GW inspiral cross section summed over all three possible inspiral pairs, denoted here by ${\sigma}_{I}$, can  be written in the more familiar astrophysical units as,
\begin{equation}
{\sigma}_{I} \approx 0.025\ \text{AU}^2 \left(\frac{v_{\infty}}{10\ \text{km\ s}^{-1}}\right)^{-2} \left(\frac{m}{M_{\rm \odot}}\right)^{12/7} \left(\frac{a_{\rm 0}}{\text{AU}}\right)^{2/7},
\label{eq:cs_GWinsp_EMC_astrounits}
\end{equation}
where we have made use of Equation \eqref{eq:cs_GWinsp_EM} with $\mathscr{F} = 6$ (see Section \ref{sec:Summary On Our Analytical Model}).

It is worth noting that the GW inspiral cross section does not scale linearly with mass $m$, as the classical Newtonian outcomes including
collisions, but instead as $m^{12/7}$. As a result, the GW cross section for, say, three $30M_{\odot}$ BHs is $\approx 200$
times larger than that for three $1.4M_{\odot}$ NSs, and not just by a factor $\approx 20$ as inferred from gravitational focusing.
In the following section we study how slight changes to the mass hierarchy introduces corrections to the equal mass case solution derived here.

\subsection{Unequal Mass Interactions}\label{sec:GW Inspirals in Unequal Mass Interactions}
To explore how sensitive our derived GW inspiral cross section is to variations in the mass ratio, we here consider the following unequal mass example,
\begin{equation}
[\text{BH}_{\rm A}, \text{BH}_{\rm B}] \leftarrow \text{BH}_{\rm B},
\label{eq:ABB}
\end{equation}
where the corresponding masses are denoted by $m_{\rm A}$ and $m_{\rm B}$, respectively. For this interaction we study the formation of
[BH$_{\rm B}$, BH$_{\rm B}$] GW inspirals.
To facilitate comparison,  we refer  to this case as {\it U} (Unequal), and {\it E} (Equal) to the
case when all three objects have  $m=m_{\rm B}$.
An example illustrating the formation of a GW inspiral for scenario {\it U}, where  $m_{\rm A}=10M_{\odot}$ and $m_{\rm B}=20M_{\odot}$, is
shown in Figure \ref{fig:GWinsp_ill1}.
Below we derive the inspiral boundary and  corresponding cross section of case {\it U} relative to {\it E},
as a function of $m_{\rm A}$. 

\subsubsection{Dependence on Mass Ratio $m_{\rm A}/m_{\rm B}$}\label{sec:Mass Ratio Dependency}

Making use of Equations \eqref{eq:e_insp} and \eqref{eq:eps_GW_EM}, the GW inspiral boundary of scenario {\it U} relative to {\it E}, takes the following form
\begin{equation}
\frac{\epsilon_{I_{\rm BB}}^{\text{(U)}}}{\epsilon_{I_{\rm BB}}^{\text{(E)}}} =  q \left(\frac{3q}{2+q}\right)^{{1}/{7}},
\label{eq:eps_GW_ABB}
\end{equation}
where we have introduced the mass ratio $q$ given by
\begin{equation}
q = \frac{m_{\rm A}}{m_{\rm B}}.
\label{eq:def_q}
\end{equation}
By the same token, the corresponding GW inspiral cross section for scenario {\it U} relative to {\it E} can be written as  
\begin{equation}
\frac{{\sigma}_{I_{\rm BB}}^{\text{(U)}}}{{\sigma}_{I_{\rm BB}}^{\text{(E)}}} = \frac{{\mathscr{I}^{\text{(U)}}}{\mathscr{F}^{\text{(U)}}}}{{\mathscr{I}}^{(E)}{\mathscr{F}}^{(E)}} \times q \left(\frac{3q}{2+q}\right)^{{1}/{7}} \times \left(\frac{2+q}{3}\right),
\label{eq:cs_GWinsp_ABB}
\end{equation}
where we have used  Equations \eqref{eq:final_sigma_Iij} and \eqref{eq:cs_GWinsp_EM}.
As can be seen, the above fraction  is composed of three different terms.
The \emph{first term} (sampling term) mainly reflects differences in the $[a',e]$ distribution and corresponding
sampling frequency of IMS binaries, where the \emph{second term} (kinematic term) relates to changes
in the inspiral time and isolation time. The \emph{third term} arises from the change in gravitational
focusing as the total mass is varied.

To explore how sensitive our derived cross section ratio in Equation \eqref{eq:cs_GWinsp_ABB}
is to changes in $q$, we now expand it to linear order in $\delta$ where $q = 1+\delta$,
\begin{equation}
\frac{{\sigma}_{I_{\rm BB}}^{\text{(U)}}}{{\sigma}_{I_{\rm BB}}^{\text{(E)}}} \approx 1 + \delta \left[\frac{30}{21} + \frac{\text{d}}{\text{d}\delta} \left(\frac{{\mathscr{I}^{\text{(U)}}}{\mathscr{F}^{\text{(U)}}}}{{\mathscr{I}}^{(E)}{\mathscr{F}}^{(E)}}\right) \bigg\rvert_{\delta = 0}  \right].
\label{eq:cs_GWinsp_ABB_EXP}
\end{equation}
As discussed in Section \ref{sec:Summary On Our Analytical Model}, we have not yet been able to write out a precise
form for the sampling term and because of this,  its derivative has been written explicitly in the  above expression.
We do have an expression for $\mathscr{I}$, however, when considering variations in $q$ it has to be consistently paired with $\mathscr{F}$.
Focusing on the remaining terms, we see that small changes in mass ratio result in small changes of order $(30/21)\delta$.
In fact, the resultant change is likely to be even smaller, as our numerical simulations indicate
that the sampling term decreases  with increasing  $\delta$.
This is explained by the fact that the heavier objects  are generally more prone to form
binaries than lighter ones \citep[e.g.][]{Sigurdsson:1993jz}. As a result, the two terms in the square parentheses from
Equation \eqref{eq:cs_GWinsp_ABB_EXP} almost cancel out, which leads to  small  differences between scenarios {\it U} and {\it E}.
We show that  this  is indeed  the case in the numerical example explored in the following section.

\subsubsection{Formation of GW Inspirals from  $m_{\rm A} = 10 M_{\rm \odot}$ and  $m_{\rm B} = 20 M_{\rm \odot}$}\label{sec:Formation of BH Binary GW Mergers}

To study a concrete example we here explore a few aspects related to the dynamical formation
of [BH(20$M_{\rm \odot}$), BH(20$M_{\rm \odot}$)] GW inspirals
for $m_{\rm A} = 10 M_{\rm \odot}$ and $m_{\rm B} = 20 M_{\rm \odot}$.
We note that $q=0.5$ is actually a rather `extreme' case from a perspective of BBH mergers forming in GCs,
as recent simulations indicate that the median mass ratio for the merging population is about $0.9$ \citep{2016PhRvD..93h4029R}. 

\begin{figure}
\centering
\includegraphics[width=\columnwidth]{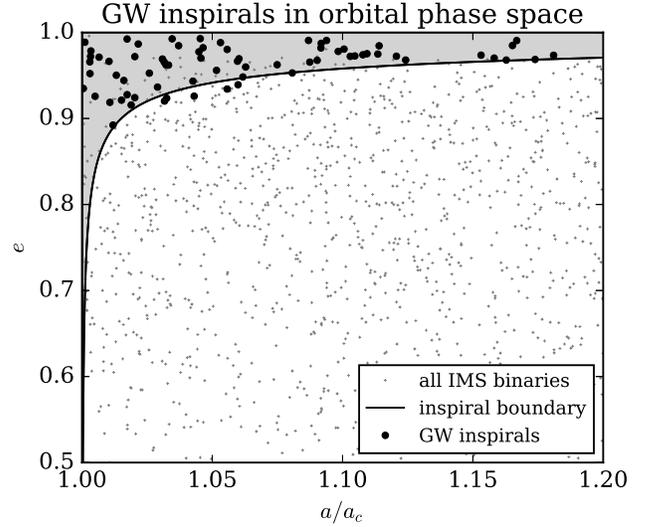}
\caption{Orbital parameter phase space [$a',e$] for all [BH(20$M_{\rm \odot}$), BH(20$M_{\rm \odot}$)]
IMS binaries formed during the binary-single interaction described
in Section \ref{sec:Formation of BH Binary GW Mergers}.
The data is based on a total of $10^{4}$ scatterings with initial SMA $a_{0} = 10^{-4}$ AU. The small value for $a_{0}$ is chosen for
illustrative purposes. The \emph{small grey dots} show [$a',e$] for all IMS binaries that are not significantly affected by
GW emission. This is in contrast to the \emph{large black dots} which show [$a',e$] for IMS binaries that after being formed undergo
a GW inspiral while still being bound to the single object.
The \emph{solid black line} shows our analytical estimate for the inspiral boundary given by Equation \eqref{eq:eps_GW_ABB}.
We see that our solution provides an accurate description of the GW inspirals obtained using our $N$-body code
for this unequal mass example.}
\label{fig:IMS_insp_ill_SIM}
\end{figure}

We first consider the [$a', e$] distribution of all [BH(20$M_{\rm \odot}$), BH(20$M_{\rm \odot}$)] IMS binaries. Results are shown in Figure \ref{fig:IMS_insp_ill_SIM}.
As seen, despite the often enormously complex pathway from initial interaction to final GW inspiral (See e.g. Figure \ref{fig:GWinsp_ill1}), we do find
excellent agreement between our full $N$-body simulations (large black dots) and our analytical solution for unequal
mass given in Equation \eqref{eq:eps_GW_ABB} (black solid line/grey area).
 Figure \ref{fig:IMS_insp_ill_SIM} also shows that the [$a',e$] distribution is indeed approximately uniform
at high eccentricity, as we assumed in Section \ref{sec:Inspiral Probability}.

\begin{figure}
\centering
\includegraphics[width=\columnwidth]{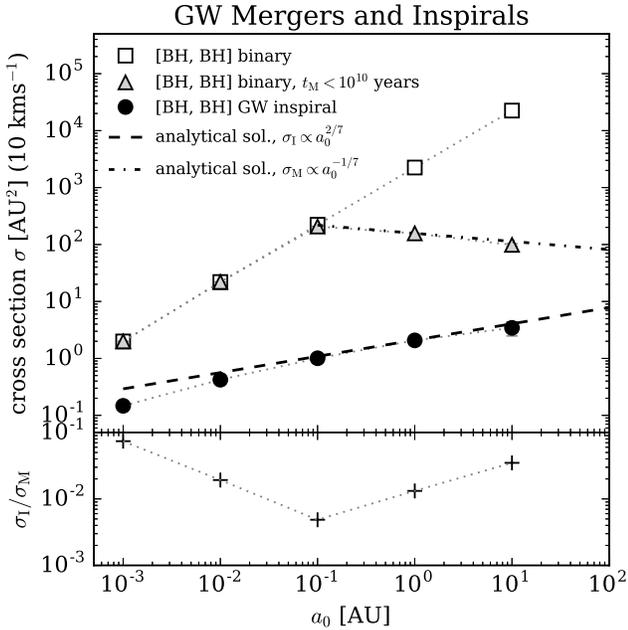}
\caption{
{\it Top panel:}
Derived cross sections as a function of initial binary SMA $a_{\rm 0}$, for
interactions between a [BH($10M_{\rm \odot}$), BH($20M_{\rm \odot}$)] binary and an incoming BH($20 M_{\rm \odot}$),
as  described in Section \ref{sec:Formation of BH Binary GW Mergers}.
\emph{Large black dots:} Cross section for [BH($20M_{\rm \odot}$), BH($20M_{\rm \odot}$)] GW inspirals.
\emph{White squares:} Cross section for an exchange interaction where the resultant binary
is [BH($20M_{\rm \odot}$), BH($20M_{\rm \odot}$)].
\emph{Grey triangles:} Cross section for an exchange interaction where the resultant binary
is [BH($20M_{\rm \odot}$), BH($20M_{\rm \odot}$)] and have a GW merger life time $t_{\rm  M}<10$ Gyrs (Hubble time).
\emph{Black dashed line:} Analytical scaling solution from Equation \eqref{eq:final_sigma_Iij} to the inspiral cross section (large black dots).
\emph{Black dashed-dotted line:} Analytical solution given by Equation \eqref{eq:Mij_tLTtau_final} to the post-interaction
binary merger cross section (grey triangles).
{\it Bottom panel:} Corresponding ratio from the top plot between the GW inspiral cross section and the post-interaction binary merger cross section.
}
\label{fig:GW_insp_102020_example}
\end{figure}

The associated GW inspiral cross section is shown in Figure \ref{fig:GW_insp_102020_example} as a function of $a_{0}$. As expected, our generic prediction from
Equation \eqref{eq:final_sigma_Iij}, which states that the inspiral cross section always scales $\propto a_{\rm 0}^{2/7}$ (for $\beta = 7/2$) in the asymptotic limit
independently of the mass hierarchy, clearly seems to hold. As a consequence, the GW inspiral cross section
increases here by more than an order of magnitude across the considered interval, and will keep increasing until
reaching the SB limit, at which it will sharply drop off \citep{2014ApJ...784...71S}.
Considering the normalization, we see that the equal mass solution from Equation \eqref{eq:cs_GWinsp_EMC_astrounits} with $m = 20M_{\odot}$
provides a rather accurate estimate (the equal mass case gives $\approx1.5\ \text{AU}^{2}$ at $a_{0} = 1$ AU), which agrees
with that describing slightly unequal mass interactions often can be done assuming the equal mass limit.

\subsection{Formation Rate of GW Inspirals}\label{sec:Formation Rate of GW Inspirals}

Having derived the cross section of GW inspirals, we are now in a position to estimate the corresponding rate.
For this, we consider the population of GW inspirals forming in the cores of GCs. As the BH mass ratios are likely to be near
unity at late times \citep{2016PhRvD..93h4029R}, we approximate individual scatterings by the equal mass limit -- an assumption
that also was shown in \cite{2017arXiv170309703S} to accurately reproduce several observables related to BBH mergers.
We take each GC to have a velocity dispersion $v_{\rm \infty}=10$ km s$^{-1}$, a core volume $V=(0.1\text{pc})^{3}$,
a total number of BHs in the core $N_{\rm BH}$ with mass $m_{\rm BH}$, and a corresponding BH binary fraction of $0.5$.
The SMA $a_{0}$ distribution is assumed to follow a Gaussian in $\log(a_{0})$ with center value $a_{\rm c}$ and standard deviation $s = 0.5$,
as has recently been shown to provide a reasonable fit to numerical simulations \citep{2016PhRvD..93h4029R}.
Adopting a GC number density of $1$ GC Mpc$^{-3}$ from \citep{2015PhRvL.115e1101R} and making use of Equation \eqref{eq:Gamma_I_gaussian}, we  find
\begin{equation}
{\Gamma_{I}} \approx 1\ \text{Gpc$^{-3}$} \text{yr}^{-1} \left(\frac{N_{\rm BH}}{80}\right)^{2} \left(\frac{m_{\rm BH}}{20M_{\odot}}\right)^{12/7} \left(\frac{a_{\rm c}}{\text{AU}}\right)^{2/7},
\label{eq:Gamma_I_GC}
\end{equation}
where ${\Gamma_{I}}$ here denotes the rate summed over all three possible inspiral pairs.
We thus conclude that  if each GC  has about 80 interacting BHs, then
about $1$ GW inspiral could be observed by LIGO per year. As GW inspirals are likely to enter the LIGO
band with non-zero eccentricity \cite[e.g.][]{2017arXiv170309703S}, we note that the possibilities for detecting GW inspirals depend highly
on constructing accurate GW templates for varying eccentricity and spin \citep[e.g.][]{2016PhRvD..94b4012H, 2016arXiv160905933H}.
Although our formalism   is well defined, our estimated  for the number of GW inspirals is  highly uncertain. This is not a unique
problem related to our model, as nearly all proposed BBH merger channels are plagued by the same large uncertainties related  to the dynamics and BH demographics of GCs \citep[e.g.][]{2017ApJ...836L..26C}.

\section{Post-Interaction Mergers and GW Inspirals}\label{sec:Inspirals Compared to Post-Interaction GW Mergers}

The vast majority of binary-single interactions end as a binary with an unbound single, also known as a \emph{fly-by} or
an \emph{exchange} outcome \citep[e.g.][]{Hut:1983js}.
Binaries formed in this way, here referred to as {\it post-interaction binaries},
will eventually merge through GW emission, however, the merger delay time
distribution is generally extremely broad with a tail that often exceeds the Hubble time.
This is in stark contrast to GW inspirals that are rare and merge relative promptly.

In this section we first calculate the cross section of
such post-interaction binary mergers. We then derive an expression for the cross section ratio
between GW inspirals and post-interaction binary mergers.
This leads us to the interesting conclusion that the cross section of GW
inspirals is {\it always} $>1\%$ of the post-interaction binary merger cross section.
The corresponding implications for BBH formation and relative rates are
studied in Section \ref{sec:Formation of Binary Black Hole Mergers}.

\subsection{Post-Interaction Binary Merger Cross Section}\label{sec:Post-Interaction GW Merger Cross Section}

We start by calculating the cross section for a post-interaction binary $[i,j]$ to merge within
a time $\tau$. This cross section, denoted here by $\sigma_{M_{\rm ij}}^{<\tau}$, can be factorized as,
\begin{equation}
\sigma_{M_{\rm ij}}^{<\tau} \approx P(BS_{\rm  ij}) P(t_{M_{\rm ij}} < \tau) \times \sigma_{\rm  CI}, 
\label{eq:Mij_tLTtau}
\end{equation}
where $BS_{\rm  ij}$ denotes an end-state composed of binary $[i,j]$ and an unbound single, $P(BS_{\rm  ij})$ is the probability for
outcome $BS_{\rm  ij}$, and $P(t_{M_{\rm ij}} < \tau)$ is the probability for that a newly formed $BS_{\rm  ij}$ binary has
a merger time $t_{M_{\rm ij}}$ that is less than $\tau$.
As we will show, the cross section $\sigma_{M_{\rm ij}}^{<\tau}$ first increases with $a_{0}$
until $a_{0}$ approaches  a characteristic value  $a_{\rm 0}^{<\tau}$, above which  $\sigma_{M_{\rm ij}}^{<\tau}$  begins to decrease with $a_{0}$.
The transitional SMA $a_{\rm 0}^{<\tau}$ is to leading order the $a_{0}$ for which all corresponding $BS_{\rm  ij}$ binaries have $t_{M_{\rm ij}} = \tau$ for eccentricity $e=0$.
For $a_{\rm 0} < a_{\rm 0}^{<\tau}$, the cross section $\sigma_{M_{\rm ij}}^{<\tau}$ is therefore simply given by $P(BS_{\rm  ij}) \times \sigma_{\rm  CI} \propto a_{0}$.
Our numerical scattering results shown
in Figure \ref{fig:GW_insp_102020_example} clearly illustrate this piecewise behavior with $a_{0}$.
In the following we estimate  $\sigma_{M_{\rm ij}}^{<\tau}$ for $a_{\rm 0} > a_{\rm 0}^{<\tau}$.

For a newly formed $BS_{\rm  ij}$ binary to merge within a time $\tau$ for a given $a_{\rm 0} > a_{\rm 0}^{<\tau}$,
its eccentricity $e$ must be larger than some characteristic value $e_{\rm \tau}$. Assuming a thermal distribution for the end-state
eccentricities \citep{Heggie:1975uy}, the probability for a $BS_{\rm  ij}$ binary to have $e>e_{\rm \tau}$ is simply
$1-{e^2_{\rm \tau}}$. From which it follows,
\begin{equation}
P(t_{M_{\rm ij}} < \tau) \approx 1-{e^2_{\rm \tau}}.
\end{equation}
Using the relation between $1-e^{2}$ and the
GW merger time of a high eccentricity binary \citep[e.g.][]{Peters:1964bc}, the cross section
$\sigma_{M_{\rm ij}}^{<\tau}$ for $a_{\rm 0} > a_{\rm 0}^{<\tau}$ can now be written as,
\begin{equation}
\sigma_{M_{\rm ij}}^{<\tau} \approx \sigma_{\rm \mathscr{R}_{\rm ij}} \times \xi,
\label{eq:Mij_tLTtau_final}
\end{equation}
where 
\begin{equation*}
\xi= \left[ \mathscr{E}^{{2}/{7}} \frac{P(BS_{\rm ij})}{\mathscr{N}\ln(a'_{\rm u})} \left( \frac{c\tau}{\pi\mathscr{R}} \frac{4\mu_{\rm ij}}{m_{\rm ij}} \sqrt{\frac{m_{\rm 1}m_{\rm 2}}{m_{\rm i}m_{\rm j}}} \right)^{{2}/{7}} \left(\frac{a_{\rm 0}}{\mathscr{R}}\right)^{-{1}/{7}} \right],
\end{equation*}
and  we have expressed  $\sigma_{M_{\rm ij}}^{<\tau}$ in a form that is similar to the inspiral cross section given in Equation \eqref{eq:final_sigma_Iij}.
Our  prediction for  $\sigma_{M_{\rm ij}}^{<\tau}$ is shown  in Figure \ref{fig:GW_insp_102020_example} as a dash-dotted line. As for the factor $\mathscr{F}$, the term $P(BS_{\rm ij})$ does not
depend on either the absolute mass scale, the initial SMA $a_{\rm 0}$, nor on any finite size effects including orbital energy losses.
However, a full analytical solution to $P(BS_{\rm ij})$ for a general mass hierarchy is not yet available. 

\subsubsection{Equal Mass Solution}

In the limit where all three objects are compact and have the same mass $m$,
we find that the post-interaction binary merger cross section summed over all three potential merger pairs,
denoted here as $\sigma_{M}^{<\tau}$, can be written in piecewise form as
\begin{equation}
\sigma_{M}^{<\tau} \approx
\begin{cases}
85.0\ \text{AU$^{2}$} \left(\frac{v_{\infty}}{10\ \text{km\ s}^{-1}}\right)^{-2} \left(\frac{m}{M_{\odot}}\right)  \left(\frac{a_{0}}{\text{AU}}\right)\\[2ex]
0.75\ \text{AU$^{2}$} \left(\frac{v_{\infty}}{10\ \text{km\ s}^{-1}}\right)^{-2} \left(\frac{m}{M_{\odot}}\right)^{{13}/{7}} \left(\frac{a_{0}}{\text{AU}}\right)^{-{1}/{7}} \left(\frac{\tau}{t_{\rm H}}\right)^{{2}/{7}}.
\end{cases}
\label{eq:PI_crosssec_EM}
\end{equation}
Here the top solution is valid  when $a_{0} < a_{\rm 0}^{<\tau}$, while the bottom solution is valid  when  $a_{0} > a_{\rm 0}^{<\tau}$.
For this equal mass case $a_{\rm 0}^{<\tau}$ is approximately given by the SMA for which the
initial target binary has a GW merger time that equals $\tau$, from which it  follows that
\begin{equation}
a_{\rm 0}^{<\tau} \approx 0.015\ \text{AU} \left(\frac{m}{M_{\odot}}\right)^{3/4} \left(\frac{\tau}{t_{\rm H}}\right)^{1/4}.
\end{equation}
If we evaluate this for $m=20M_{\odot}$ and $\tau=t_{\rm H}$ we find $a_{\rm 0}^{<\tau} \approx 0.15$ AU.
This partly explains why most of the BBH mergers formed in numerical simulation pile up
near this SMA \citep{2016PhRvD..93h4029R}.

\subsection{GW Inspirals Relative to Post-Interaction GW Mergers}\label{sec:GW Inspirals Relative to Post-Interaction GW Mergers}

\begin{figure}
\centering
\includegraphics[width=\columnwidth]{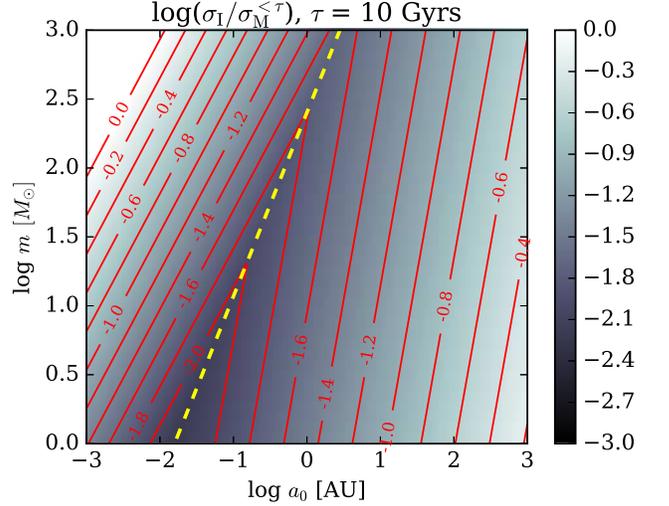}
\caption{The {\it contour values} show the log of the ratio between the GW inspiral cross section and the post-interaction binary
merger cross section in the equal mass case for $\tau = 10$ Gyrs, as a function of initial SMA $a_{0}$ and mass $m$.
The {\it yellow dashed} line shows the combination of $a_{0}$ and $m$ for which the corresponding GW merger life time of the initial target binary is $=\tau$.
As argued in Section \ref{sec:GW Inspirals Relative to Post-Interaction GW Mergers}, the yellow line also traces the minimum
value of ${\sigma_{I}}/{\sigma_{M}^{<\tau}}$. As a result, for $m \gtrsim 10M_{\odot}$
the ratio will always be larger than $\approx 0.01$ independently of $a_{0}$, which suggests that the GW inspiral cross section will {\it always} be at least $1\%$ of the
the post-interaction binary merger cross section. This has profound implications for estimating the number of high eccentricity binary BH sources observable by
Advanced LIGO, as further described in Section \ref{sec:Formation of Binary Black Hole Mergers}.
}
\label{fig:GWinsp_over_GWmerg}
\end{figure}

The ratio between the GW inspiral cross section and the post-interaction GW merger cross section
can be simply  written in the equal mass case as
\begin{equation}
\frac{\sigma_{I}}{\sigma_{M}^{<\tau}} \approx
\begin{cases}
2.9\cdot10^{-4} \left(\frac{m}{M_{\odot}}\right)^{{5}/{7}}  \left(\frac{a_{0}}{\text{AU}}\right)^{-{5}/{7}},  \\[2ex]
3.3\cdot10^{-2} \left(\frac{m}{M_{\odot}}\right)^{-{1}/{7}} \left(\frac{a_{0}}{\text{AU}}\right)^{{3}/{7}} \left(\frac{\tau}{t_{\rm H}}\right)^{-{2}/{7}},
\end{cases}
\label{eq:sigmaI_over_sigmaM_EM}
\end{equation}
where we have used Equations \eqref{eq:cs_GWinsp_EMC_astrounits} and \eqref{eq:PI_crosssec_EM}, respectively.
The top expression here again applies  for $a_{0} < a_{\rm 0}^{<\tau}$, while the bottom  is for $a_{0} > a_{\rm 0}^{<\tau}$.
The value of $\log({\sigma_{I}}/{\sigma_{M}^{<\tau}})$ is illustrated in Figure \ref{fig:GWinsp_over_GWmerg},
as a function of initial SMA $a_{0}$ and mass $m$ for $\tau = t_{\rm H}$. 

As given by Equation \eqref{eq:sigmaI_over_sigmaM_EM}, and also seen in
both Figures \ref{fig:GW_insp_102020_example} and \ref{fig:GWinsp_over_GWmerg},
the ratio ${\sigma_{I}}/{\sigma_{M}^{<\tau}}$ behaves opposite to the scaling of ${\sigma_{M}^{<\tau}}$, as it first
decreases with $a_{0}$  until $a_{0} \approx a_{\rm 0}^{<\tau}$, after which it increases with $a_{0}$.
This piecewise scaling interestingly implies that the ratio ${\sigma_{I}}/{\sigma_{M}^{<\tau}}$
takes its minimum value at which the initial SMA $a_{0} \approx a_{\rm 0}^{<\tau}$. Evaluating the ratio at $a_{0}  = a_{\rm 0}^{<\tau}$,
therefore lead us to the following inequality
\begin{equation}
\frac{\sigma_{I}}{\sigma_{M}^{<\tau}} \gtrsim 0.01 \left(\frac{m}{10M_{\odot}}\right)^{5/28} \left(\frac{\tau}{t_{\rm H}}\right)^{-5/28},
\label{eq:min_sigmaI_over_sigmaM}
\end{equation}
which states that the GW inspiral cross section will {\it always} be at least $1\%$ of the post-interaction binary merger
cross section for interacting compact objects of similar mass $m \gtrsim 10M_{\odot}$.

\section{Formation of Binary Black Hole Mergers}\label{sec:Formation of Binary Black Hole Mergers}

Having developed  a clear understanding of GW inspirals and post-interaction binary mergers, we are now finally in a position to
study their relative contributions to the number of dynamically formed  BBHs.
Our goal in this section is to examine GW inspirals arising from dynamical interactions in a dense stellar system
like a GC. Given the uncertainties surrounding the presence and role of BHs in the evolution of GCs and other dense stellar clusters,
we restrict ourselves to two idealized scenarios. In the first {\it Early-Burst Scenario}, we study the formation of BBH mergers that form from a population of binaries and
singles that interact only over a brief time interval. In the second {\it Steady-State Scenario} we study the
formation of BBH mergers from an interacting,  steady state population. Results from the two scenarios are given below.

\subsection{Early-Burst Scenario}

Let us  first assume a population of binaries and singles undergoing interactions over a short time interval at some initial time $t_{0}$.
We speculate that this could represent the dynamical environment of an early collapse associated
with the formation of a dense
stellar system, such as a GC \citep[e.g.][]{2013MNRAS.429.2298M, 2014ApJ...794....7M, 2014MNRAS.441.3703Z, 2015MNRAS.451.4086S, 2016MNRAS.463.2443K}.
For this scenario we study the cumulative number of BBH mergers forming through the GW inspiral channel, $N_{I_{\rm ij}}$, relative to that of the
post-interaction binary merger channel, $N_{M_{\rm ij}}$, as a function of time $\tau = t - t_{0}$. We note  that only $N_{M_{\rm ij}}$ is
time dependent because of its delay time distribution.
For this scenario it is straight forward to show that
\begin{equation}
\frac{N_{I_{\rm ij}}}{N_{M_{\rm ij}}} \approx \frac{\sigma_{I_{\rm ij}}}{\sigma_{M_{\rm ij}}^{<\tau}}.
\label{eq:NIij_NMij_EBS}
\end{equation}
The equal mass solution to the {\it Early-Burst Scenario}  is therefore simply given by Equation \eqref{eq:sigmaI_over_sigmaM_EM}.
This  implies that the inequality derived in Equation \eqref{eq:min_sigmaI_over_sigmaM} can be applied to this case, from which we conclude that
the cumulative number of GW inspirals in the equal mass limit for this scenario will always be at least $1\%$
of the total number of BBH mergers for $m \gtrsim 10M_{\odot}$.

The solution to Equation \eqref{eq:NIij_NMij_EBS} also shows that the relative number of GW inspirals scales
$\propto \tau^{-2/7}$, which interestingly suggests that the first BBH mergers in the Universe were likely to be GW inspirals. More generally, the solution implies that there exists a finite time interval within
which the fraction of BBH mergers is dominated by GW inspirals and not by standard post-interaction binary mergers.
In the equal mass case, we can solve for this characteristic time interval, denoted here as $\tau_{\rm IM}$, by setting
${N_{I}} = {N_{M}}$, from which we find
\begin{equation}
\tau_{\rm IM} \approx 2.0\times 10^{6}\ \text{yrs} \left(\frac{m}{20M_{\odot}}\right)^{-1/2} \left(\frac{a_{0}}{25\text{AU}}\right)^{3/2}.
\end{equation}
The normalization of the initial SMA $a_{0}$ is here set to $25\ \text{AU}$, which is $\approx 0.1$ times the HB value for
$v_{\infty} = 10\ \text{km\ s}^{-1}$ and $m = 20M_{\odot}$. Despite large uncertainties in what the relevant values
of $a_{0}$ and $m$ should on average be, this simple estimate  indicates that a characteristic time interval of order $\approx10^{6}\ \text{years}$ is not unrealistic.
We note this is similar to the dynamical time of a typical GC.

These results indicates  that GW inspirals  might play a role in the formation of a dense stellar system at
early times. If we denote the ratio between the number of GW inspirals to the number of binary-single interactions by $f_{\rm IB}$,
one finds in the equal mass limit  that 
\begin{equation}
a_{0} \approx 0.15\ \text{AU} \left(\frac{f_{\rm IB}}{0.01}\right)^{-7/5} \left(\frac{m}{20M_{\odot}}\right).
\end{equation}
This indicates that for $m \approx 20M_{\odot}$, the initial SMA must be $\lesssim 0.15$ AU for GW inspirals to
form in at least $1\%$  of all binary-single interactions. GW inspirals are indeed more likely to
contribute to the dynamical evolution of a dense  stellar system than post-interaction binary mergers.
This is because GW inspirals form and evolve in bound resonant states, which is in contrast to post-interaction binary mergers that always
receive a dynamical kick prior to merger. This   kick is often high  enough to unbound them  from the dense stellar system \citep[e.g.][]{2016PhRvD..93h4029R}.
This could have  interesting dynamical  consequences, and numerical simulations do in fact show
an indication for this to be the case \citep{2004ApJ...616..221G, 2006ApJ...640..156G}, yet, more work is needed before firm conclusions can be drawn. 

\subsection{Steady-State Scenario}

As a second representative example, we study the formation of BBH mergers in a
GC assuming its distribution of binaries and
singles remains constant in time.
We imagine this steady-state scenario to
approximately describe the late time evolution and formation of BBH mergers
in a typical GC. Below we first calculate the cumulative number of BBH mergers
and then derive relative rates,
from which we find that $\gtrsim 1\%$ of the present day rate of BBH mergers assembled through binary-single interactions
is  likely to originate from the GW inspiral channel.
Finally, in the last section, we estimate the fraction of GW inspirals that is likely to
appear in the LIGO band with a particular high eccentricity.

\subsubsection{Cumulative Number of BBH Mergers}

Here we derive the cumulative number of GW inspirals and post-interaction binary mergers
forming as a function of time $\tau = t-t_{0}$, consistently taking into account ongoing interactions
as well as post-interaction binary merger delay time distributions. Assuming $a_{0} > a_{\rm 0}^{<\tau}$,
one can show that the time dependent cumulative number of post-interaction binary mergers for this scenario can be written  as
\begin{equation}
N_{M_{\rm ij}} \propto \int_{0}^{\tau} \sigma_{M_{\rm ij}}^{<\tau'} d\tau' = \frac{7}{9} \tau \sigma_{M_{\rm ij}}^{<\tau},
\end{equation}
and the cumulative number of GW inspirals as,
\begin{equation}
N_{I_{\rm ij}} \propto \int_{0}^{\tau} \sigma_{I_{\rm ij}} d\tau' = \tau \sigma_{I_{\rm ij}},
\end{equation}
where we have assumed that the average inspiral time $ \ll \tau$. The cumulative number of GW inspirals relative to that of the post-interaction binary mergers
at time $\tau$, is thus  given by
\begin{equation}
\frac{N_{I_{\rm ij}}}{N_{M_{\rm ij}}} \approx \frac{9}{7} \frac{\sigma_{I_{\rm ij}}}{\sigma_{M_{\rm ij}}^{<\tau}}.
\label{eq:NI_over_NM}
\end{equation}
We  note that  this estimate and the one derived for the early-burst scenario
differ by only a factor of $9/7$. This  indicates that the relative number of GW inspirals forming through the binary-single channel is
not strongly dependent on the encounter  history. In the equal mass case, it seems robust to conclude that
$>1\%$ of all BBHs forming through the binary-single channel are likely to be in form of GW inspirals.

\subsubsection{Relative Rate of GW inspirals}

The rate of GW inspirals relative to that of the post-interaction binary mergers evaluated at time $t \gg t_{0}$, can be shown to take the
following  form
\begin{equation}
\frac{\Gamma_{I_{\rm ij}}}{\Gamma_{M_{\rm ij}}} \approx \frac{\sigma_{I_{\rm ij}}}{\sigma_{M_{\rm ij}}^{<t}},
\end{equation}
Assuming the equal mass limit, we can combine this estimate
with  Equation \eqref{eq:min_sigmaI_over_sigmaM} to obtain 
\begin{equation}
\frac{\Gamma_{I}}{\Gamma_{M}} \gtrsim 0.01 \left(\frac{m}{10M_{\odot}}\right)^{5/28}.
\end{equation}
This again gives further support to our  conclusion that the rate of GW inspirals will  be
$\gtrsim 1\%$ of the total rate for  $m \gtrsim 10M_{\odot}$.

From an astrophysical perspective, GW inspirals are of particular interest because of their high eccentricity at formation.
However, only a fraction of the GW inspirals will actually appear in the LIGO band
with the same high eccentricity they had at formation as a result of circulation during inspiral \citep[e.g.][]{2017arXiv170309703S}.
In the section below we derive this fraction and its present day relative rate.

\subsubsection{GW Inspirals Forming in the LIGO Band}

One can show that for a BBH to appear in the LIGO band with its initial peak eccentricity, it must necessarily {\it form}
within the band \citep[e.g.][]{2014ApJ...784...71S}. The fraction of GW inspirals at a high eccentricity is thus given to leading order by the fraction
that forms within  the observable LIGO band. 

By the use of the GW peak frequency fitting formula from \cite{Wen:2003bu},
one can solve for  the required  initial orbital parameters of an inspiraling BBH so that
 its GW peak frequency is above a certain threshold $f_{\rm GW}$. This  provides us with  the following relation
\begin{equation}
r_{f_{\rm GW}} \approx 10^{-5}\ \text{AU} \left(\frac{f_{\rm GW}}{10\text{Hz}}\right)^{-2/3} \left(\frac{m}{20M_{\odot}}\right)^{1/3},
\end{equation}
where we have assumed the high eccentricity limit and $r_{f_{\rm GW}}$ is the pericenter distance at which the corresponding GW peak frequency equals $f_{\rm GW}$. 
If the BBH pericenter distance $r_{\rm p} < r_{f_{\rm GW}}$ then the GW peak frequency will be $\gtrsim f_{\rm GW}$.
As $r_{f_{\rm GW}}$ is a fixed distance, we can now conclude that the cross section for a BBH
to form with GW peak frequency $\gtrsim f_{\rm GW}$ is to leading order given by  Equation \eqref{eq:cs_Rij} with
${R}_{\rm C} = r_{f_{\rm GW}}$, and not by the GW inspiral cross section. From this it follows directly that
\begin{equation}
{\sigma}_{f_{\rm GW}} \approx 0.4\ \text{AU}^2 \left(\frac{v_{\infty}}{10\ \text{km\ s}^{-1}}\right)^{-2} \left(\frac{m}{20M_{\rm \odot}}\right)^{4/3} \left(\frac{f_{\rm GW}}{10\text{Hz}}\right)^{-2/3},
\end{equation}
where ${\sigma}_{f_{\rm GW}}$ is the cross section for a BBH to form with an associated GW peak frequency $\gtrsim f_{\rm GW}$ after a binary-single interaction.
Comparing this with the GW inspiral cross section from Equation \eqref{eq:cs_GWinsp_EMC_astrounits}, one finds the relative rate in the equal mass limit to be of order
\begin{equation}
\frac{{\Gamma}_{f_{\rm GW}}}{{\Gamma}_{I}} \approx 0.2 \left(\frac{m}{20M_{\rm \odot}}\right)^{-8/21} \left(\frac{f_{\rm GW}}{10\text{Hz}}\right)^{-2/3} \left(\frac{a_{0}}{0.1\text{AU}}\right)^{-2/7}.
\end{equation}
This illustrates  that for the chosen normalizations only $\approx  20\%$ of the GW inspirals will have a GW peak frequency $f_{\rm GW}>10\text{Hz}$ at
the time of their formation.
In other words, only $\approx 20\%$ of the GW inspirals will appear in the LIGO band with their initial high eccentricity, whereas the
remaining $\approx 80\%$ will undergo notable circularization before being  detected.
Although this seems slightly discouraging in terms of observable rates, we do note that the vast majority of GW inspirals
still make it into the LIGO band at $10\text{Hz}$ with significant eccentricity (typically $>0.1$). This was shown
by \cite{2017arXiv170309703S}, who highlighted  that if the BBH merger rate
is dominated by the GC population, then GW inspirals are likely to dominate the population of eccentric BBH mergers
observable by LIGO.

\section{Conclusions}\label{sec:Conclusions}

The few-body problem with terms correcting for finite sizes, tides, and GR effects, is usually studied
using  $N$-body codes \citep[e.g.][]{Mardling:2001dl, 2016arXiv160909114S}. However, these 
generally provide very limited physical insight into the problem.
In this paper we have explored analytically the effect from including such correction terms in three-body interactions, which
have provided  us with a wealth of new insight into this problem.
Our study focus on the population of two-body captures that form through tidal and GW energy losses in
binary-single interactions -- a population that we have been referring  to as {\it inspirals}. In order to provide a clear physical framework,  we have first developed an
analytical formalism  for calculating the inspiral cross section and corresponding rate.
We then applied it to explore the formation of GW inspirals,
as we found this population to be highly interesting both dynamically and observationally.
Our of findings are summarized below.

\subsection{Finite Sizes and Energy Loss Terms}
As illustrated in \cite{2016arXiv160909114S}, the main effect from including finite sizes and orbital energy loss term corrections
in binary-single interactions, is an increase in the number of mergers, that either form through standard collisions (by the finite size term) or
inspirals (by tides and GW emission). This increase is not related to an overall change in the dynamics,
as the addition of the energy loss  terms does not significantly modify the number of classical exchange and fly-by outcomes.
Instead, the correction terms lead to occasional highly impulsive orbital energy losses that
then result in either a collision or an inspiral.
As the capture distance is always larger than the corresponding physical size, the inclusion of orbital energy loss terms
will always lead to an increase in the number of mergers. Our formalism shows how this effective size, and thereby the number of mergers,
varies with the initial orbital energy of the three-body system.

\subsection{Inspiral Cross Section and Rates}
We found that the cross section of inspirals increases independently of the mass hierarchy as $a_{0}^{1/\beta}$,
when the considered orbital energy loss term scales with the pericenter distance as $r_{\rm p}^{-\beta}$. This provides an accurate description for GW emission,
but it is only approximate for tides \citep[e.g.][]{2016arXiv160909114S}.
The collision cross section is found  to be independent of $a_{0}$ and, as such,  we concluded
that the number of mergers relative to that of collisions always increases with $a_{0}$. 
The relative importance of inspirals is thus partly set by the HB limit; systems with low velocity dispersion are therefore likely to show the largest
effect from orbital energy loss term corrections.
The corresponding inspiral rate can be written in closed form for a few relevant cases as a result of our analytical solution.
As an example, we showed that for a dynamical system where the target binaries follow a Gaussian in $\log(a_{0})$ with standard deviation $s$ and central value $a_{\rm c}$,
the Gaussian assumption only plays a role when $s>\beta$. However, this condition is rarely met in
typical GC systems\footnote{$s$ is often about $\approx 0.5$ as found by \cite{2016PhRvD..93h4029R}, where $\beta$ is always $>3.5$ as discussed in \cite{2016arXiv160909114S}},
and the rate is, to leading order, simply proportional to the inspiral cross section evaluated at $a_{\rm c}$.

\subsection{Formation and Importance of GW Inspirals}
The inclusion of GW energy loss terms leads to the formation of GW inspirals. To study this population,
we started by investigating the relative contribution of GW inspirals and collisions  to the number of GW mergers formed during  binary-single interactions. Using our analytical framework, we found that GW inspirals generally are $\approx  100$ times more likely to form than standard collisions for encounters involving 
BHs and NSs. A consistent inclusion of GW energy loss terms in $N$-body codes is therefore crucial for accurately estimating 
rates of compact object mergers. We note here that the public code {\it Fewbody} \citep{Fregeau:2004fj, 2012ascl.soft08011F} does not include such terms in
its original version, which indeed has led to a long list of studies that significantly underestimate
the number of especially highly eccentric GW mergers. This was recently illustrated by \cite{2017arXiv170309703S}.

\subsection{Cross Section and Rate of GW Inspirals}
While our formalism directly shows that  the GW inspiral cross section varies as $a_{0}^{2/7}$, it
is not clear how sensitive it is to changes in the initial mass hierarchy.
To explore this, we have performed a controlled analytical experiment, where only one of the masses
was varied away from its equal mass value. In this case we found that
the change in GW inspiral cross section is only weekly dependent  on the fractional mass change.
This  led us to conclude that the equal mass case actually seems to provide a fairly good description
of BBH mergers assembled in a typical GC, as these recently have been found to
have a mass ratio close to unity \citep{2016PhRvD..93h4029R}. This was also illustrated in \cite{2017arXiv170309703S}.
In this approximation, we have showed that the rate of GW inspirals is of order $\Gamma_{I} \approx 1\ \text{Gpc}^{-3}\ \text{yr}^{-1}$ for a typical GC population, if
the number of BHs in each cluster is about $\approx 80$. As the number of BHs in GCs is largely unknown, this rate estimate is of course
associated with a large uncertainty. For this reason, we  mainly explore relative rates in this paper.

\subsection{Formation of Post-Interaction GW Mergers}
The majority of  GW mergers do not form during the interaction as  inspirals,
but instead as post-interaction binary  mergers. To facilitate  comparison, we have calculated
the cross section of such post-interaction binary mergers and  found it to have a piecewise scaling with $a_{0}$. The cross section is shown to  first increases as $\propto a_{0}$ until a
characteristic value $a_{\rm 0}^{<\tau}$, after which it decreases as $\propto a_{0}^{-1/7}$. We thus conclude that the
post-interaction binary merger cross section reaches its maximum value when $a_{0} = a_{\rm 0}^{<\tau}$.
In the equal mass case, the value of $a_{\rm 0}^{<\tau}$ is about the SMA for which the target binary has a GW merger time $\tau$.

\subsection{GW Inspirals and Post-Interaction GW Mergers}
Relative rates can be determined much more accurately than absolute ones. For this reason, we have explored the cross section of  inspirals relative to that of post-interaction binary mergers.
We found that the relative number of GW inspirals takes its minimum value when $a_{0} = a_{\rm 0}^{<\tau}$,
which led us to the profound  conclusion that $>1\%$ of all the BBH mergers assembled through binary-single interactions
will be GW inspirals. As a result, if post-interaction binary mergers are as frequent as recently reported by \cite{2016PhRvD..93h4029R}, GW inspirals are  expected  to be
observed by Advanced LIGO.

\subsection{Formation History of GW Inspirals}
To gain further insight into the role of GW energy loss term corrections, we have calculated the time dependent number of GW inspirals and post-interaction binary mergers
for two scenarios.
In the first scenario, we have assumed that binaries and singles interact over a short period. In this case we have found that GW inspirals likely dominate the total GW merger rate within the first $\approx 10^{6}$ years. This interestingly implies that the first mergers in the Universe are likely to be inspirals.
In the second scenario, we have explored  the rates assuming steady state.
For this case, we have found that the present day rate of GW inspirals relative to that of post-interaction binary mergers is approximately given by the
ratio of their cross sections evaluated at $\tau = t_{\rm H}$. As a result, GW inspirals are expected to constitute at least $1\%$ of the present day rate of
BBHs mergers.
We also  noted that GW inspirals are more likely to remain in the GC after their formation than standard post-interaction binary mergers,
which could have  interesting dynamical consequences \citep[e.g.][]{2006ApJ...640..156G}.

\subsection{Highly Eccentric GW Mergers Observable by LIGO}
Although the majority of GW inspirals form with notable eccentricity, many might still experience significant circularization
before being observable \citep[e.g.][]{2017arXiv170309703S}. As a result, for a BBH to appear in the LIGO band
with an extremely high eccentricity, it must necessarily form within the LIGO band. Using a simple prescription for
calculating the GW peak frequency $f_{\rm GW}$, we have showed that a fixed value for $f_{\rm GW}$
corresponds to a fixed pericenter distance $r_{\rm p}(f_{\rm GW})$. The cross section
of GW inspirals that form with a GW frequency, say, $>10$Hz, is therefore to leading order given by the collision cross section with $R_{\rm C} = r_{\rm p}(10\text{Hz})$.
Making use of this, we  derived that $\approx 20\%$ of all GW inspirals are expected to form within the LIGO band and thus appear
with a particular high eccentricity. We note that this estimate  agrees well with the one derived using $N$-body simulations \citep{2017arXiv170309703S}.
Although $\approx 80\%$ all inspirals are not expected to form within the LIGO band, we still find that the majority of GW inspirals will enter the band with
an eccentricity $>0.1$ \citep{2017arXiv170309703S}.  If the BBH merger rate is dominated by the dynamically assembled population,
then about $1\%$ of the observable rate will have an eccentricity $>0.1$. 

The results presented here  clearly illustrate that GW inspirals are not just a rare curiosity resulting from including
GR corrections, but constitute a population with highly interesting dynamical and observational consequences. This study
should motivate further work on understanding the role of GR corrections in few-body interactions as well as in full $N$-body GC calculations.
As GW inspirals generally merge with notable eccentricity, they are likely to play a key role in differentiating between different BBH merger channels;
a test  that should become possible  with Advanced  LIGO.

\acknowledgments{
It is a pleasure to thank C. L. Rodriguez, N. Leigh and T. Ilan for helpful discussions.
Support for this work was provided by  the David and Lucile Packard Foundation, UCMEXUS (CN-12-578), the Danish National Research Foundation and  NASA through an Einstein Postdoctoral Fellowship grant number PF4-150127, awarded
by the Chandra X-ray Center, which is operated by the
Smithsonian Astrophysical Observatory for NASA under
contract NAS8-03060.
}

\bibliographystyle{apj}


\end{document}